\documentclass[journal]{IEEEtran}    

\usepackage{times}
\usepackage{amssymb}
\usepackage{amsmath}
\usepackage{amsthm}
\usepackage{amsfonts}
\usepackage{txfonts}
\usepackage{graphicx}
\usepackage{algorithm}
\usepackage{algorithmic}
\usepackage{subfig}

\newtheorem{theorem}{Theorem}
\newtheorem{lemma}{Lemma}

\newtheorem{definition}{Definition}

\begin{document}
\title{FlexAuc: Serving Dynamic Demands in a Spectrum Trading Market with Flexible Auction}

\author{Xiaojun~Feng,~\IEEEmembership{Student~Member,~IEEE,}
        Peng~Lin,
        Qian~Zhang,~\IEEEmembership{Fellow,~IEEE,}
\thanks{X. Feng, P. Lin and Q. Zhang are with the Department
of Computer Science and Engineering, Hong Kong University of Science and Technology,
Hong Kong.  e-mail: \{xfeng, linpeng, qianzh\}@cse.ust.hk.}
}

\maketitle

\begin{abstract}
In secondary spectrum trading markets, auctions are widely used by spectrum holders (SHs) to redistribute their unused channels to secondary wireless service providers (WSPs).
As sellers, the SHs design proper auction schemes to stimulate more participants and maximize the revenue from the auction.
As buyers, the WSPs determine the bidding strategies in the auction to better serve their end users.

In this paper, we consider a three-layered spectrum trading market consisting of the SH, the WSPs and the end users. We jointly study the strategies of the three parties. The SH determines the auction scheme and spectrum supplies to optimize its revenue. The WSPs have flexible bidding strategies in terms of both demands and valuations considering the strategies of the end users.
We design FlexAuc, a novel auction mechanism for this market to enable dynamic supplies and demands in the auction.
We prove theoretically that FlexAuc not only maximizes the social welfare but also preserves other nice properties such as truthfulness and computational tractability.

\end{abstract}


\section{Introduction}

Auctions are usually applied for spectrum redistribution to increase efficiency.
In the spectrum trading markets, major wireless service providers (WSPs) purchase spectrum through auctions organized by the spectrum holders (SH).
It is essential for the SH to determine the auction schemes and for the WSPs to determine proper bidding strategies in the auction to optimize their revenues.

Here we have two key observations.
First, the number of channels on sale in the auction have an significant impact on the revenue of the SH. Given a fixed total bandwidth, a finer channelization provides more supplies, but induce higher costs in terms of larger guard bands and higher complexity.
However, no existing works in the literature have consider this strategies of SH. The most related one is  \cite{wang2012towards}. In \cite{wang2012towards}, Wang et al. considers the resource segmentation between auction and another pricing scheme for a cloud computing market. The supplies of cloud instances in the auction can be dynamic. But they fail to consider different capacities of one instance which is equivalent to different channelization in the spectrum auctions.

Second, the WSPs' bidding strategies relate tightly to the service provisions to the end users. Both the WSPs' demands and valuations can be flexible in the auction due to their end users' heterogeneous demands and willingness to pay.
However, existing works failed to enable the WSPs' flexible bidding strategies.
There are no suitable existing auction schemes that are able to promote flexible biddings with computational efficiency.
Most works rely on over-simplified assumptions or induce heavy overheads.
Some of them (\cite{zhou2008ebay, jia2009revenue, zhou2009trust}) assume that one buyer can claim for at most one channel. Some others \cite{yang2011truthful}\cite{peng2013groupon} assume that a buyer can submit bids for multiple channels but win at most one.
These assumptions indeed limit the scope of applications especially when a buyer is willing to win more than one channels.
Although combinatorial auctions (e.g. \cite{dong2012combinatorial}) can meet the requirements of flexible auction, it induces very heavy computational overheads.

Considering these flexibilities, it is therefore important for the SH to determine proper channelization of the unused spectrum and to design auction schemes to motivate the buyers to reveal their truthful demands and valuations.

\begin{figure}[t]
\centering
\includegraphics[width=3.0in]{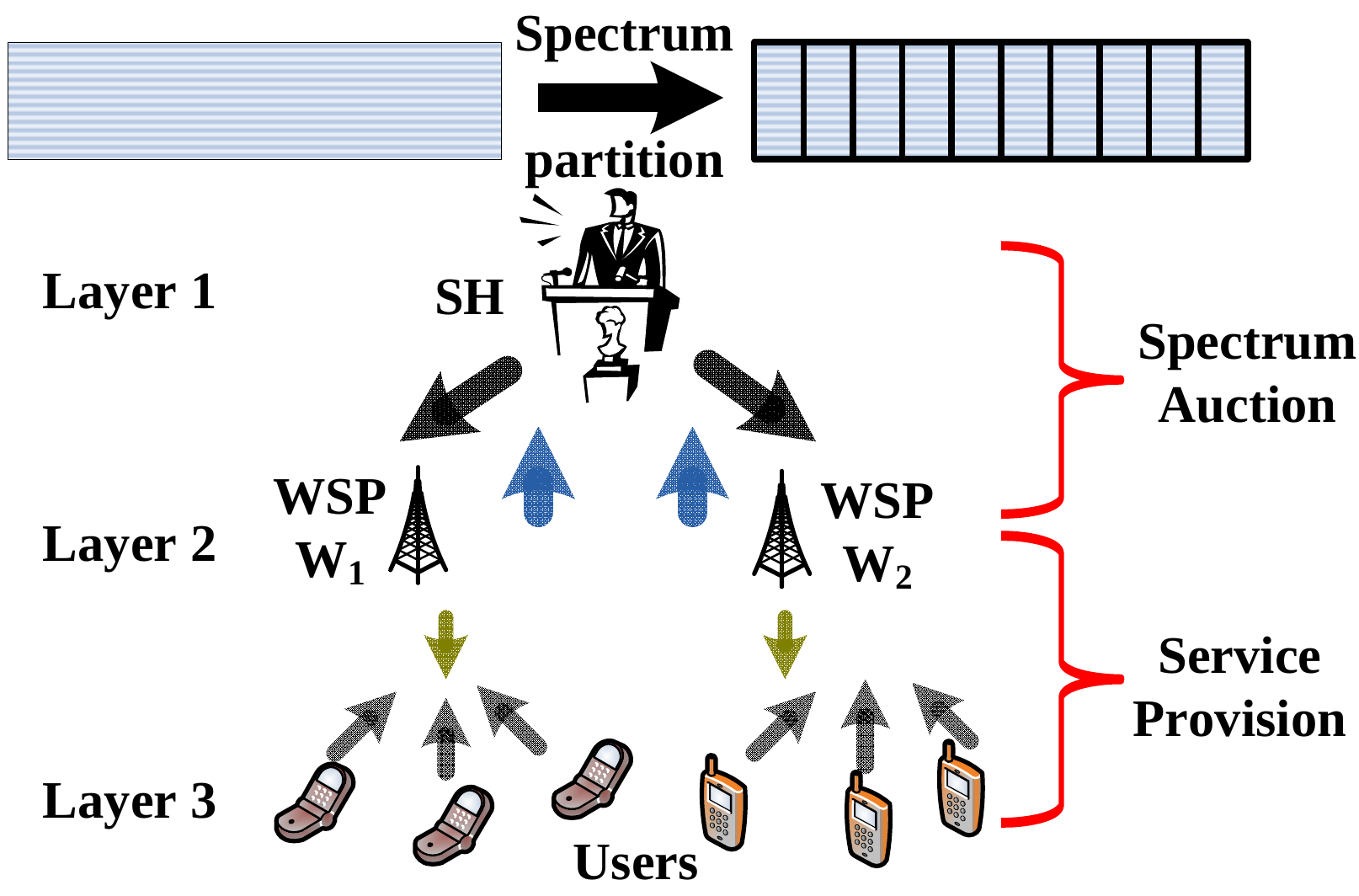}
\caption{A three-layered network scenario}
\label{fig:sys}
\end{figure}

In this paper, we aim to address the two problems mentioned above.
Specifically, we consider a three-layered spectrum trading scenario shown in Fig. \ref{fig:sys} consisting of the SH, the WSPs and the end users.
The SH has a fixed sized unused spectrum to be redistributed to the WSPs via an secondary spectrum auction.
As a seller, the SH partitions the spectrum into multiple channels with equal bandwidth and designs the auction mechanism. The SH can determine the size of a channel.
As buyers, the WSPs decide their bidding strategies in the auction, which is flexible in term of both demands and bids.
With the channels won in the auction, the WSPs serve and charge their end users.
All the three parties involved in the scenario are rational and trying to optimize their own utility.

In this paper, we propose a new secondary spectrum trading framework to solve the challenges in this problem. The key component in our solution framework is a novel auction mechanism called FlexAuc (\textbf{Flex}ible \textbf{Auc}tion).
In FlexAuc, the SH announces the total number of channels and the bandwidth of each channel for sale. The channels are indifferent for the same WSP. But may be valuated differently across WSPs. Then the WSPs each submit a series of bids to represent their willingness to pay for the channels in an ordinal order. For example, if an WSP would like to pay the first won channel with payment $b_1$ and the second won channel with $b_2$, the bids would be $\{b_1, b_2\}$.
In this way, both the demand and the valuations of WSPs can be flexibly expressed in the auction.
We propose two new payment mechanisms to allow the SH to charge the winners in the auction.
We prove in the paper that the auction scheme with different payment mechanisms are all not only truthful but can also maximize social welfare.

Along with FlexAuc, we also jointly consider the strategies of the SH in terms of the channelization and the service provision between the WSPs and their end users.
The SH chooses the best spectrum partitioning scheme to optimize its revenue.
The WSPs decide their services prices according to the results from the auction.

In summary, the key conclusions and contributions of this work are summarized as
follows.
\begin{enumerate}
\item As far as we know, this paper is the first to study the flexible auction mechanism in the secondary spectrum trading market, where the WSPs can flexibly decide the best quantity of channels to buy and the best bidding values.

\item We design FlexAuc, with two novel payment mechanisms to serve flexible demands in secondary spectrum auctions. We prove that FlexAuc is not only truthful but also maximize social welfare.

\item We jointly study the channelization problem for the SH and the service provision problem for the WSPs and derive optimize strategies to maximize their revenues.
\end{enumerate}

The rest of this paper is organized as follows. We summarize the
research literature in Section \ref{sec:relatedwork}. The detailed
system model, concept definitions and design objectives are
described in Section \ref{sec:systemmodel}. In Section
\ref{sec:flexauc}, we discuss the spectrum partition, and elaborate
auction mechanism, bidding behaviors and the demand response as well
as pricing mechanism. We analyze the economic properties in Section
\ref{sec:economic_prop}, specifically truthfulness, efficiency and
time complexity. Section \ref{sec:num} provides our numerical
results to verify our design objectives and make comparison with
previous works to show advantages of FlexAuc. We conclude this paper
in Section \ref{sec:conclusion}.

\section{Related Work}\label{sec:relatedwork}

We provide a summary of state-of-the-art auction mechanisms in this section.

Most of the state-of-the-art mechanisms assume that buyers can claim at most one channel.
These mechanisms cannot satisfy the flexible demands from buyers.
Single-seller multi-buyer auction with homogeneous channels has been studied extensively.
VERITAS in \cite{zhou2008ebay} allowed users to buy channels based on their demands and spectrum owner to maximize revenue with spectrum reuse.
In \cite{jia2009revenue} the authors proposed a VCG auction to maximize the expected revenue of the seller and a suboptimal auction to reduce the complexity for practical purpose.
Double auction mechanisms are studied for the multi-seller multi-buyer case. McAfee mechanism \cite{mcafee1992dominant} was proposed for trading homogeneous items in double auction.
Many follow-up works has been done since then \cite{zhou2009trust,wu2012strategy,yang2011truthful,feng2012tahes}.

One recent mechanism for heterogeneous demands is designed for the cloud services \cite{zhang2013framework}.
Unlike our work, in \cite{zhang2013framework}, each bidder has only uniform unit valuations for any amount of demands.
Another mechanism that enables the flexible demand is combinatorial auction \cite{dong2012combinatorial}\cite{zhang2013strategy}.
In this model, each buyer can submit a bid to flexibly claim time-channel combinations.
The major concern for combinatorial auction is that in general cases, the problem is NP-hard.
It means that the mechanism is not suitable for periodic auctions with too many buyers and channels.

Furthermore, no existing works on spectrum auction considers the strategies of spectrum sellers in terms of channelization. The most related one is \cite{wang2012towards}, which is focused on a cloud computing market. In \cite{wang2012towards}, Wang et al. study the resource segmentation between a periodic auction scheme and a pay-as-you-go scheme. The supplies of cloud instances in the auction can be dynamically allocated. Unlike \cite{wang2012towards} where one cloud instance has a fixed capacity, in this paper, the SH can vary the channel bandwidth and create different supplies in the auction.
In a preliminary version of this paper \cite{peng2014Flexauc}, we also omit the channelization strategy of the SH.

In summary, none of the existing works provided a feasible auction scheme to enable flexible demand with polynomial time complexity.
Most polynomial time auction models do not support flexible demand. Combinatorial auction supports flexible demand, but they are not computational efficient. FlexAuc preserves good properties of both types. Furthermore, this paper is the first to consider the channelization strategies of the SH for secondary spectrum auctions.

\section{Preliminaries and Problem Formulation}\label{sec:systemmodel}

In this section, we first describe our general scenario, then the definition of the strategies, utility functions and concepts.

\subsection{System Model}

We consider a scenario of a macrocell with one SH denoted as $\mathcal{S}$ and $N$ WSPs denoted as $W=\{\mathcal{W}_1, \mathcal{W}_2, \cdots, \mathcal{W}_N\}$.
The SH has a spectrum block of total bandwidth size $B_0$ to sell.
The bandwidth is partitioned into $C$ channels, each with the same bandwidth $B$.
To avoid interference between adjacent channels, we assume a guard band of size $b_0$ should be placed between every pair of adjacent channels.
Therefore, there are $C-1$ guard bands for $C$ channels. It is easy to check that the following relationship holds:
\begin{equation}
B=\frac{B_0+b_0}{C}-b_0.
\label{eqn:guardband}
\end{equation}

Each WSP deploys infrastructures within the coverage of the same macrocell as the SH.
The WSPs obtained the bandwidth from the SH via an auction organized by the SH to serve its end users.
A WSP $\mathcal{W}_i$'s bid in the auction is defined as a vector consisting of $C$ bids: $\mathcal{B}^i=\{b^i_1, b^i_2, \cdots, b^i_C\}$.
It means that $\mathcal{W}_i$ is willing to pay $b^i_1$ for the first winning channel, $b^i_2$ for the second winning channel, and so on.

The WSPs also charges their end users with certain service fees. We assume the price for the usage of unit bandwidth of $\mathcal{W}_i$ is $p_i$. Due to the diversified capacity of the WSPs in terms of transmission powers and service types, the service qualities of the WSPs can also be different. We use $\alpha_i$ to denote the evaluation of the unit data rate from $\mathcal{W}_i$ to describe its service quality.

We assume that there are $N_i$ end users subscribing to the WSP $\mathcal{W}_i$ ($1 \leq i \leq N$). We denote the $j$-th end user of $\mathcal{W}_i$ ($1\leq j \leq N_i$) as $\mathcal{U}^i_j$.
The end users of an WSP can decide how much bandwidth to subscribe to maximize their own utility.
Suppose the bandwidth strategy of $\mathcal{U}^i_j$ is $w^i_j$, we assume that it can achieve a data rate of
$
R^i_j=w^i_j\ln\left(1+\frac{P_iH_j}{n_0w^i_j}\right),
$
where $P_i$ is the transmission power level of base stations of $\mathcal{W}_i$; $H_j$ is the path loss factor; and $n_0$ is the power density of thermal noise.
Let $g^i_j=\frac{P_iH_j}{n_0}$ denote the factor related to the signal-to-noise ratio for $\mathcal{U}^i_j$. We can rewrite $R^i_j$ as:
\begin{equation}
R^i_j=w^i_j\ln\left(1+\frac{g^i_j}{w^i_j}\right).
\end{equation}

\subsection{Spectrum Trading and Pricing Procedure}

\begin{figure}[t]
\centering
\includegraphics[width=3.0in]{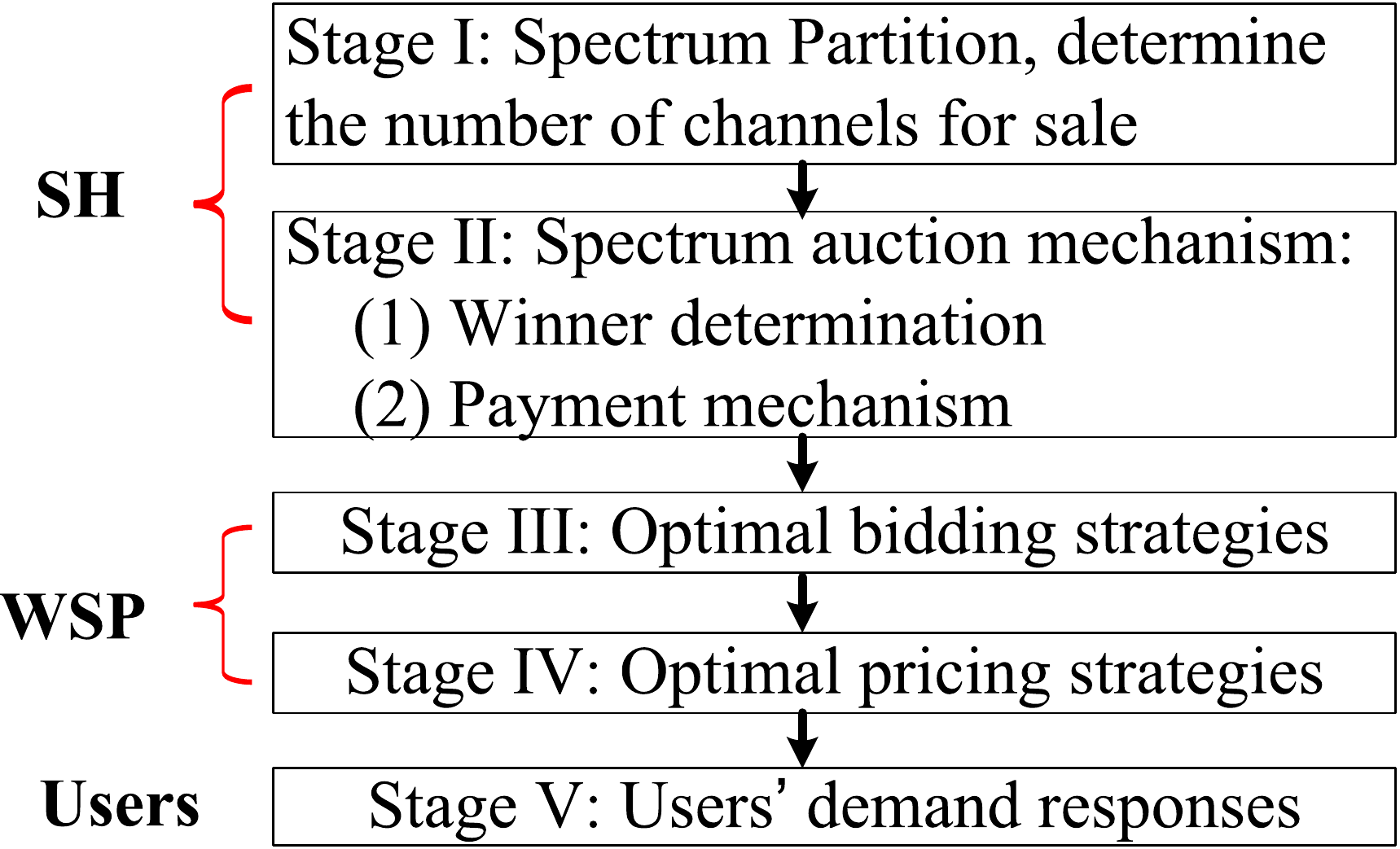}
\caption{A four-staged analysis framework.}
\label{fig:procedure}
\end{figure}

The spectrum trading and pricing procedure involves the SH, the WSPs and their end users. There are five steps in the procedure shown in Fig. \ref{fig:procedure}.

The SH first decides the channelization scheme for the available spectrum, i.e determines the number of channels to be distributed in the auction and the auction schemes in terms of winner determination and payment mechanisms in Stage I and II respectively. Being informed about the number of channels for sale and the auction mechanism, the WSPs then determines their bidding strategies in stage III. When the auction is complete, the WSPs decide pricing schemes for the end users based on the winning channels obtained in stage IV. Finally in stage V, the end users decide how much demand to subscribe from the WSPs given the prices.

\subsection{Utility Functions}

Based on the system model and the trading procedure, we can define the utility functions for the SH, the WSPs and the end users.

\subsubsection{Spectrum Holder}

The utility of the spectrum holder is defined as the revenue from the auction. In the spectrum auction, the SH determines the winner of each channel and the prices. Suppose the number of channels won by $\mathcal{W}_i$ is $K_i$, each with price $c^i_j, 1\leq j \leq K_i$. The utility of the SH can be written as:
\begin{equation}
U_S=\sum_{i=1}^{N} \sum_{j=1}^{K_i}c^i_j.
\label{formula:sh_uti}
\end{equation}

\subsubsection{WSPs}

The utility of the WSPs are defined as the difference between the payment from the end users and the prices paid in the auction.
For $\mathcal{W}_i$, the total bandwidth demand from the end users is $\sum_{j=1}^{N_i}w^i_j$. It will try to satisfy these demands. However, $\mathcal{W}_i$ may not obtain exactly the same bandwidth as the total demands from the auction.
Follow the same notations used above, we can define the utility for $\mathcal{W}_i$ as:
\begin{equation}
U_{i}=p_i \cdot \min\left\{\sum_{j=1}^{N_i}w^i_j, K_iB\right\}-\sum_{j=1}^{K_i}c^i_j. \label{formula:wsps_uti}
\end{equation}
We can see that if $\mathcal{W}_i$ obtains more bandwidth from the auction, it can only charge the end users according to their total demands.

\subsubsection{End Users}

Considering the service price and evaluation of unit data rate, we define the utility of $\mathcal{U}^i_j$ as the difference between its evaluation of the unit data rate and the total payment of the obtained bandwidth from $\mathcal{W}_i$. Note that $\alpha_i$ is the evaluation of the unit data rate from $\mathcal{W}_i$.
\begin{equation}
U^i_j=\alpha_{i} R^i_j - p_iw^i_j
     = \alpha_{i} w^i_j \ln\left(1+\frac{g^i_j}{w^i_j}\right)-p_iw^i_j.
\label{formula:sus_uti}
\end{equation}

\begin{table}[tp]
\renewcommand{\arraystretch}{1.3}
\caption{Key notations in this paper.}
\label{tab:Notations}
\centering
\begin{tabular}{|c|p{3.0in}|}
  \hline
  $\mathcal{S}$ & The spectrum holder in our scenario \\
  \hline
  $U_S$ & The utility of $\mathcal{S}$ \\
  \hline
  $N$ & The number of WSPs in our scenario \\
  \hline
  $B_0$ & The total bandwidth for sale from the SH \\
  \hline
  $C$ & The number of channels for sale \\
  \hline
  $B$ & The bandwidth of each channel \\
  \hline
  $b_0$ & The bandwidth of guard band between channels \\
  \hline
  $W$ & The set of all WSPs \\
  \hline
  $\mathcal{W}_i$ & The WSP with index $i$ \\
  \hline
  $p_i$ & The price for unit bandwidth provided by $\mathcal{W}_i$ \\
  \hline
  $p^*_i$ & The optimal price for $\mathcal{W}_i$ \\
  \hline
  $\alpha_i$ & The evaluation for unit data rate provided by $\mathcal{W}_i$ \\
  \hline
  $N_i$ & The number of end users subscripted with $\mathcal{W}_i$ \\
  \hline
  $\mathcal{B}^i$ & The bidding vector of $\mathcal{W}_i$ in the auction \\
  \hline
  $b^i_j$ & The bid for the $j$-th winning channel of $\mathcal{W}_i$ \\
  \hline
  $b^s_k$ & The $k$-th highest bids in the auction submitted by the WSPs \\
  \hline
  $\widetilde{b}^i_j$ & The true valuation for the $j$-th winning channel by $\mathcal{W}_i$ \\
  \hline
  $c^i_j$ & The price for the $j$-th winning channel of $\mathcal{W}_i$ in the auction \\
  \hline
  $K_i$ & The number of channels won by $\mathcal{W}_i$ in the auction \\
  \hline
  $U_i$ & The utility of $\mathcal{W}_i$ \\
  \hline
  $\mathcal{U}^i_j$ & The $j$-th end user of $\mathcal{W}_i$ \\
  \hline
  $w^i_j$ & The bandwidth demand of $\mathcal{U}^i_j$ \\
  \hline
  $w^{*i}_j$ & The optimal bandwidth demand of $\mathcal{U}^i_j$ \\
  \hline
  $g^i_j$ & A factor related to the signal-to-noise ratio of $\mathcal{U}^i_j$ \\
  \hline
  $G_i$ & The sum of $g^i_j$ of all end users of $\mathcal{W}_i$ \\
  \hline
  $U^i_j$ & The utility of $\mathcal{U}^i_j$ \\
  \hline
\end{tabular}
\end{table}

\subsection{Definition of Concepts}

In this part, we clarify some concepts we will use in this paper and the economic properties we would like to achieve.

\begin{definition}
Dominant Strategy: a dominant strategy of a player is the one that
maximize its utility regardless of what other players' strategies
are. Mathematically, if $x_i$ is player $i$'s strategy, for any
$x'_i \neq x_i$, and any strategy profile of others $x_{-i}$, we
have $U_i(x_i,x_{-i}) \geq U_i(x'_i,x_{-i})$. If the inequality
always holds, $x_i$ is a strongly dominant one. Otherwise, $x_i$ is
a weakly dominant one.
\end{definition}

\begin{definition}
Truthfulness: an auction is truthful if any player's true evaluation
is its dominant strategy.
\end{definition}
It means that given other players' strategy profile and the auction
rules fixed, a player cannot improve its utility by submitting any
bid that is different from its true bid (a vector of its true
evaluation).

When designing auction mechanisms, it is crucial to make them
truthful. It is the most critical property and has been well
accepted in the research literature \cite{zhou2008ebay}.

\begin{definition}
Individual Rationality: an auction is individual rational if no
buyer is charged more than its bid and no seller is paid less than
its ask. It guarantees the validness of the auction result.
\end{definition}

\begin{definition}
Social Efficiency: an auction is social efficient if the aggregate
of all players' utilities is optimized. It shares a common meaning
with Social Welfare. Here we consider the aggregate utility of the
SH and the WSPs, denoted by $S=\sum_{i=1}^N U_i+U_{S}$. The users'
utilities are not considered in Social Welfare because there is no
direct relationship between the users and the auction.
\end{definition}

\section{Spectrum Trading Framework Design}\label{sec:flexauc}

Based on the system model and spectrum trading and pricing procedure, in this section, we leverage backward induction to analyze the strategies of the SH, the WSPs and the end users.
In this paper, we assume all of them are rational and trying to maximize their own utility.

We first analyze the pricing decision of the WSPs in stage IV based on the possible reactions from the end users in stage V. The optimal bidding strategies in terms of number of channels to purchase and individual evaluations is determined in stage III based on possible revenue from the pricing. The auction scheme in stage II will be decide by the SH considering only the economic properties to achieve. We observe that the WSPs can have flexible demands and evaluation. To enable such flexibilities and extract higher revenue, we will present the design of the FlexAuc (Flexible Auction) mechanism for the SH. Based on the auction scheme and possible outcomes, in Stage I, the SH determines optimal channelization schemes on the spectrum.

\subsection{Users' Demand Strategies}

Given the price $p_i$ announced by $\mathcal{W}_i$, in Stage V, end users optimize their utility by selecting the optimal demand strategies.
For the ease of analysis, we make an approximation:
\begin{equation}
\label{eqn:approximation}
U^i_{j} \approx \alpha_{i} w^i_j \ln\frac{g^i_j}{w^i_j}-p_iw^i_j.
\end{equation}
Its first and second order derivatives are
\begin{equation*}
\frac{\partial U^i_j}{\partial w^i_j} = \alpha_{i} \left(\ln
\frac{g^i_j}{w^i_j}-1\right)-p_i.
\end{equation*}
and
\begin{equation*}
\frac{\partial^2 U^i_j}{\partial {w^i_j}^2}
=-\frac{\alpha_{i}}{w^i_j}<0.
\end{equation*}
So the optimal demand strategy is
\begin{equation}
w^{*i}_j = g^i_j e^{-1-\frac{p_i}{\alpha_{i}}}.
\label{formula:su_optdemand}
\end{equation}
The error introduced by approximation in (\ref{eqn:approximation}) decreases with $p_i$.
Generally, the error is very small in our simulation with the ITU and COST models \cite{recommendation19971225}.


\subsection{WSPs' Pricing Strategies}

In stage IV, $\mathcal{W}_i$ decides the optimal price $p_i$ based on the auction result and evaluation of users' demands.

To find the optimal price $p^*_i$ that maximizes Eq. (\ref{formula:wsps_uti}), given the number of winning channel $K_i$, we need to discuss two cases.

\textbf{Case (1):} if $K_iB>\sum_{j=1}^{N_i}w^{*i}_j$, Eq. (\ref{formula:wsps_uti}) can be simplified as
\begin{equation}
p_i \cdot
\sum_{j=1}^{N_i}w^{*i}_j-\sum_{j=1}^{K_i}c^i_{j},
\end{equation}
which has the only variable $p_i$.
The unique solution $p^*_i=\alpha_i$ can be obtained.
Physically, it means that if $\mathcal{W}_i$ has purchased abundant bandwidth, the optimal price that brings highest revenue is $p^*_i=\alpha_i$. Though part of its spectrum is unused, it is nonprofitable to offer a reduced price to stimulate more spectrum demand.

\textbf{Case (2):} if $K_iB \leq \sum_{j=1}^{N_i}w^{*i}_j$, the optimal $p^*_i \geq \alpha_i$ because $p_i \sum_{j=1}^{N_i}w^{*i}_j-\sum_{j=1}^{K_i}c^i_{j}$ is monotonously decreasing in the region $p_i \in [1, \infty)$.
Therefore, the highest utility is achieved when $\sum_{j=1}^{N_i}w^{*i}_j=K_iB$.
Let $G_i=\sum_{j=1}^{N_i}g^i_j$.
We can obtain $p^*_i=\alpha_i\left(\ln\frac{G_i}{K_iB}-1\right)$, where $G_ie^{-1-\frac{p_i}{\alpha_i}}=\sum_{j=1}^{N_i}w^{*i}_j$ which is the aggregated demand of $\mathcal{W}_i$'s users. In this case, if $\mathcal{W}_i$ purchases more bandwidth, it can lower the price and increase the utility.

To make a summary, the optimal price for $K_i$ channel is
\begin{equation}
p^*_i(K_i)=\left\{
\begin{aligned} &\alpha_i && \textrm{if}
\hspace{0.15cm} K_iB > G_ie^{-2}, \\
&\alpha_i\left(\ln \frac{G_i}{K_iB}-1\right) && \textrm{Otherwise.}
\end{aligned} \right. \label{formula:optimalprice}
\end{equation}

\subsection{WSPs' Bidding Strategies}

For Stage III, we analyze WSPs' bidding strategies in the auction.
Since we take truthfulness and efficiency as the most desirable properties of an auction design, the WSPs are aware that the auction design by the SH is truthful and efficient. Therefore, the WSPs bid with their truthful evaluations of the channels.

In our problem, both the quantity and evaluation of channels are dynamic and flexible.
In such case, the challenge is how to determine the best bidding strategies in terms of quantity and evaluation. We propose to decide the best strategies by the pricing scheme and users' demand responses.

Let us start with a simple case.
Suppose $\mathcal{W}_i$ gets only one channel in the auction. Its utility can be calculated by Eq.
(\ref{formula:sus_uti}) as $U_i(K_i=1)$. When $\mathcal{W}_i$ considers the bidding strategy, it cannot predict the number of winning channels $K_i$ and the payments $\{c^i_{j}\}$. A well-designed auction mechanism guarantees that the true value is the best strategy, which simplifies the strategy making process.

First, we need to elaborate the true value in this problem. In
previous auction works, the true values exist in the mind of buyers
and sellers in advance. If a buyer is asked to pay the true value to
win the object, the utility is zero. In this paper, we propose that
the true value should be decided according to particular scenarios.
We also define the true value as the one that makes a buyer's
utility zero given that it selects the best strategies in the later
stages. We use $\widetilde{b}^i_{j}$ to denote WSP's true value for its $j$-th won channel.
So we get $\mathcal{W}_i$'s true value for its first channel as:
\begin{equation}
\widetilde{b}^i_{1}=p^*_i(1)\cdot\min\left\{\sum_{j=1}^{N_i}w^{*i}_j(1), B \right\}.
\label{formula:1c}
\end{equation}
$p^*_i(k)$ and $w^{*i}_j(k)$ stand for the optimal pricing strategy
and demand strategy given $\mathcal{W}_i$ has $k$ channels. Eq.
(\ref{formula:1c}) is also the marginal benefit that the first
channel can bring to $\mathcal{W}_i$. If $\mathcal{W}_i$ pays more than this amount to
get it, $\mathcal{W}_i$ definitely makes a loss.

Similarly, considering the marginal benefit of each additional
channel for $\mathcal{W}_i$, the true value for $\mathcal{W}_i$'s $k$-th channel ($2 \leq
k \leq C$) will be
\begin{equation}
\begin{aligned}
\widetilde{b}^i_{k}=& p^*_i(k)\cdot\min\left\{\sum_{j=1}^{N_i}w^{*i}_j(k), kB \right\} \\
&-p^*_i(k-1)\cdot\min\left\{\sum_{j=1}^{N_i}w^{*i}_j(k-1), (k-1)B \right\}.
\label{formula:kc} \end{aligned}
\end{equation}

By intuition, the optimal bids for any $\mathcal{W}_i$ are decreasing
as the marginal benefit of the first several channels is higher than
the latter ones. We can derive this relationship from Eq.
(\ref{formula:su_optdemand})(\ref{formula:optimalprice})(\ref{formula:1c})(\ref{formula:kc})
to verify its correctness. We have the theorem:

\begin{theorem}
\label{th:bid}
Any $\mathcal{W}_i$'s bidding structure presents marginal decreasing property.
Mathematically,
$
\widetilde{b}^i_{1} \geq \widetilde{b}^i_{2} \geq \cdots \geq \widetilde{b}^i_{C}.
$
\end{theorem}

The proof is provided in the Appendix.

In fact this property facilitates the algorithm design for
FlexAuc, such that we can obtain linear-time algorithms.

\begin{algorithm}[tp]
\caption{Flexible Auction: winner determination} \label{alg1}
\begin{algorithmic}[1]
\STATE Find the $C$ largest bids from $\{b^i_{j}\}, (i=1,\cdots, N, j=1,\cdots, C)$, with a max heap.
\STATE Let $\{b_{k}^{s}\}, k=1,\cdots, C$ be the sorted array of the $C$ largest bids in descending order.
\STATE $K_i=0$, for all $i=1,\cdots, N$.
\FOR {$k=1$ to $C$}
\IF {$b_k^s$ is a bid submitted by $\mathcal{W}_i$}
\STATE $K_i=K_i+1$
\ENDIF
\ENDFOR
\STATE return ($\mathcal{W}_i$ wins $K_i$ channels, $i=1,\cdots, N$).
\end{algorithmic}
\end{algorithm}

\subsection{Auction Design}

Now we analyze the auction design in stage II.
The goal in auction design is two-fold. First, the auction mechanism
should be truthful. Therefore, $\mathcal{W}_i$ bidding with its true values is
the dominant strategy. Second, the auction should provide
efficiency, which means the channels should be allocated to those
bidders who evaluate them most. Therefore, SH's revenue can be
increased.

The auction is a sealed-bid auction with one seller (SH) and
multiple buyers (WSPs). The auction procedure consists of two parts:
the winner determination (channel allocation) and payment mechanism.
As the channels are identical, the channel allocation result is
presented in the form of the number of winning channels $\{K_i\}$.
The payment mechanism can be flexible. In this paper, we consider
three different payment mechanisms.
The first one is the well-known VCG mechanism \cite{vickrey1961counterspeculation, clarke1971multipart, groves1973incentives}.
The second one is a modified version of the uniform pricing mechanism which preserves truthfulness for multi-unit demands.
Besides, we also design a partial uniform pricing mechanism which achieves higher revenue than the other two schemes.

\subsubsection{Winner determination}

The auction is a standard auction such that the $C$ bids with
highest values are selected as winning bids:
$
\sum_{i=1}^{N}K_i=C.
$
This can be easily achieved by selecting the largest $C$ bids from
all $NC$ bids submitted by the WSPs, which is done by building a maximum heap (Algorithm \ref{alg1}).
Note that the SH does not need to sort all the bids. He needs to know only the largest $C$ bids to announce them as the winning bids.

\begin{algorithm}[tp]
\caption{Flexible Auction: payment mechanism (VCG)} \label{alg2}
\begin{algorithmic}[1]
\STATE Double the size of $\{b_{k}^{s}\}$ to $2C$.
\STATE Find the $2C$ largest bids from $\{b^i_{j}\}, (i=1,\cdots, N, j=1,\cdots, C)$, with a max heap.
\STATE Store the sorted array of the $2C$ largest bids in descending order in $\{b_{k}^{s}\}, k=1,\cdots, 2C$.
\FOR {$i=1$ to $N$}
\IF {$K_i>0$}
\STATE $losingBid=0$
\STATE $j=C+1$
\STATE $totalPayment(i)=0$
\WHILE {$losingBid < K_i$}
\IF {$b_j^s$ is not submitted by $\mathcal{W}_i$}
\STATE $totalPayment(i)=totalPayment(i)+b_j^s$
\STATE $losingBid=losingBid+1$
\ENDIF
\STATE $j=j+1$
\ENDWHILE
\ENDIF
\ENDFOR
\STATE return ($\mathcal{W}_i$ pays $totalPayment(i)$, $i=1,\cdots, N$).
\end{algorithmic}
\end{algorithm}

\subsubsection{Payment mechanism}

The payment mechanism is relatively independent of the previous
winner determination part. Here we introduce three possible payment
mechanisms, namely: (i) the VCG mechanism, (ii) the modified uniform pricing
mechanism, and (iii) the partial uniform pricing mechanism. The algorithms
for payment mechanisms are following the notations in Algorithm
\ref{alg1}.

\textbf{The VCG Mechanism:}
The idea of VCG mechanism is highly abstract and can be applied to
universal cases, independent of the form of bidding structure, items
to be sold, and so on. With VCG mechanism, $\mathcal{W}_i$'s payment is
determined by the externality he exerts on other competing WSPs. In
this auction, the externality is the sum of $K_i$ highest losing bids submitted by other WSPs'.
The VCG mechanism is designed as Algorithm
\ref{alg2}. Still the same, the SH needs to know the highest $2C$ bids
only. Lines $1$ to $3$ do this job. Because even in the extreme case
where one WSP wins all $C$ channels, its externality is the
following $C$ largest bids. That means knowledge of the largest $2C$
bids is sufficient.

\textbf{The Uniform Pricing Mechanism:}
The general idea of the uniform pricing is to charge each channel the same price
as the channels are identical. It increases the buyers' acceptance
since there is no price discrimination. The uniform price charged is also
called the market clearing price.
However, in general, uniform pricing is not truthful for multi-unit demand \cite{ausubel2002demand}.
To guarantee the truthfulness, the clearing price should be selected independent of the winners' bids.
Here we introduce a modification of the traditional uniform pricing scheme.
Instead of choosing the highest losing bids or the lowest winning bids as the clearing price, we use the highest bids from the bidder who loses all its bids.
This modified version of uniform pricing only works under the condition that $C<N$.
That is because only when $C<N$, we are guaranteed to be able to find a WSP who does not win a channel at all.
In the rest of this paper, we mean ``uniform pricing'' by this modified version.
The modified uniform pricing algorithm is given by Algorithm \ref{alg3}.
In Algorithm \ref{alg3}, we find a WSP who has not won any channel in the auction via line 4. Then its highest bids would be charged for the winners as the unit price for any single channel (line 5, 8).

\begin{algorithm}[tp]
\caption{Flexible Auction: payment mechanism (Uniform Pricing)}
\label{alg3}
\begin{algorithmic}[1]
\STATE (Assert $C < N$)
\STATE $maxLoserBid=0$ // Store the highest bid from a bidder who does not win any channel \label{alg3_1}
\FOR {$i=1$ to $N$}
\IF {$b^i_{1} < b_{C}^{s}$ and $b^i_{1} > maxLoserBid$}
\STATE $maxLosrBid=b^i_{1}$
\ENDIF
\ENDFOR
\label{alg3_2}
\STATE return ($\mathcal{W}_i$ pays $totalPayment(i)=maxLosrBid\times K_i$, $i=1,\cdots, N$).
\end{algorithmic}
\end{algorithm}

\begin{algorithm}[tp]
\caption{Flexible Auction: payment mechanism (Partial Uniform Pricing)} \label{alg4}
\begin{algorithmic}[1]
\STATE // Find the largest losing bids of each bidder and stores in an array $maxLoserBid$
\FOR {$i=1$ to $N$}
\IF {$K_i < C$}
\STATE $maxLoserBid(i)=b^i_{K_i+1}$
\ELSE
\STATE $maxLoserBid(i)=0$
\ENDIF
\ENDFOR
\STATE Find the two largest one from $maxLoserBid$, denoted as $m1$ and $m2$, ($m1 > m2$).
\FOR {$i=1$ to $N$}
\IF {$maxLoserBid(i) == m1$} \label{alg4_1}
\STATE $totalPayment(i)=m2\times K_i$
\ELSE
\STATE $totalPayment(i)=m1\times K_i$
\ENDIF
\ENDFOR
\STATE return ($\mathcal{W}_i$ pays $totalPayment(i)$, $i=1,\cdots, N$).
\end{algorithmic}
\end{algorithm}

\textbf{The Partial Uniform Pricing Mechanism:}
Motivated by the previous two mechanisms, we design the partial
uniform pricing, which preserves both their advantages. By partial
uniform pricing, a winning WSP pays for each of its channels the
same amount of money. Its unit price is determined by others'
highest losing bid. But different WSPs can have different unit
prices. There are two advantages of partial uniform pricing compared with uniform pricing.
On one hand, this mechanism will generate revenue for the SH no
less than that of VCG mechanism as stated in the following Lemma:
\begin{lemma}
\label{lemma:pricing}
In any case, the revenue generated by the Partial Uniform Pricing Mechanism is larger or equal than that generated by the VCG Pricing Mechanism or the Uniform Pricing Mechanism.
\end{lemma}
We omit the formal proof here because of its simplicity. An intuitive explanation is as follows. By partial uniform pricing mechanism, $\mathcal{W}_i$ pays the amount of
$K_i$ multiples others' highest losing bid. By VCG mechanism, $\mathcal{W}_i$
pays the amount of the sum of others' highest $K_i$ bids. Also, these payment should be no lower than the highest bid submitted by a WSP who wins no channels (the unit price in the uniform pricing mechanism).

On the other hand, this mechanism does not require the condition $C<N$ since we always
have a losing bid as $\mathcal{W}_i$'s clearing price. The partial
uniform pricing mechanism is shown in Algorithm \ref{alg4}.
In Algorithm \ref{alg4}, $maxLoserBid(i)$ is the highest bid from $\mathcal{W}_i$ that does not win in the auction. For any WSP, either $m1$ or $m2$ must be the highest losing bid submitted from another WSP. In line \ref{alg4_1}, we are avoiding the case that the highest losing bid is the one submitted by $\mathcal{W}_i$, which may result in untruthful bidding.

As a summary, we can actually combine any of the three payment mechanisms with the standard
winner determination part to get a complete auction rule. The
auction is truthful with any of the three mechanisms. The auction
also maximizes social welfare. These two properties will be proved
later.

\subsection{Spectrum Partition}

In stage I, the SH decides the channelization scheme to determine the number of channels for sale in the auction.
On one hand, from Eq. (\ref{eqn:guardband}), we can see that with smaller $C$, the overhead of guard band would be smaller.
On the other hand, A small $C$ means that the SH sells bigger piece of channels to fewer WSPs. Considering that the WSPs' marginal evaluations on spectrum
bandwidth are decreasing, it is better for the SH to further divide
the big spectrum piece to smaller ones and sell them to more WSPs.
Therefore, the SH needs to find the optimal channel partitioning scheme to maximize its revenue.

To choose optimal channelization scheme, there are two challenges.
First, how can the SH estimates the bids in the auction? The flexibility of channel demands and valuations from the WSPs make this task complicated.
Second, there are three different payment choices in the auction, how can the SH determine the optimal channelization despite the difference in the payment mechanisms? We solve these challenges via the following steps.

\subsubsection{Bid Estimation}

We can derive the relationship between the revenue of the SH and the number of channels.
From Eq. (\ref{formula:optimalprice}) and Eq. (\ref{formula:kc}), we can see that $\widetilde{b}^i_{k}$ (the bids from $\mathcal{W}_i$) is a piece-wise function. However, when $k>\frac{G_ie^{-2}}{B}$, $p^*_i=\alpha_i$, therefore, $\widetilde{b}^i_{k}=0$. A bid of zero is meaningless in the auction, therefore, we can only consider the bids when they are positive. Suppose $k<\frac{G_ie^{-2}}{B}$, substitute Eq. (\ref{formula:optimalprice}) into Eq. (\ref{formula:kc}), we have:
\begin{equation}
\widetilde{b}^i_{k}=\alpha_iB\left(\ln\frac{G_ih(k)}{B}-1\right),
\label{eqn:bid}
\end{equation}
where $h(k)=\frac{(k-1)^{k-1}}{k^k}$.

Here $G_i$ and $\alpha_i$ are parameters relates to individual WSPs. The SH normally has no ways to obtain their exact values. However, in practice, based on the public information of the WSPs such as their annual report or market surveys, the SH can have an estimation of these values. For example, if the distribution of $G_i$ and $\alpha_i$ are known, the SH can leverage statistic method to estimate the bids \cite{dyspan14feng}. For simplicity, in this paper, we assume that the SHs has the estimations of $G^E_i$ and $\alpha^E_i$ for $G_i$ and $\alpha_i$ respectively. With these assumption, substitute Eq. (\ref{eqn:guardband}) into Eq. (\ref{eqn:bid}), we have:
\begin{equation}
\widetilde{b}^i_{k}=\alpha^E_i\left(\frac{B_0+b_0}{C}-b_0\right)
                    \left(\ln\frac{G^E_ih(k)C}{B_0-(c-1)b_0}-1\right).
\label{eqn:bid_channel}
\end{equation}

\subsubsection{A Uniform Indicator}

Despite the difference of the three payment mechanisms, we can use a uniform indicator to guide the channelization decision to maximize $U_S$. We denote the $k$-th highest bid submitted by the WSPs in the auction as $b^s_k$. The following relationship holds:
\begin{lemma}
\label{lemma:indicator}
If there are $C$ channels and more than one WSPs in the auction, $C\times b^s_{C+1}$ is a tight upper bound for the revenue of the SH in all the three payment mechanisms.
\end{lemma}
We prove the Lemma in the Appendix.
Since $C\times b^s_{C+1}$ is a tight upper bound for all the three payment mechanisms, we define $U^I_S$ as the indicator for $U_S$:
\begin{equation}
\label{eqn:indicator}
U^I_S=C\times b^s_{C+1}.
\end{equation}

\subsubsection{Optimal Channel Partitioning}

The optimization problem for the SH can be expresses as:
\begin{eqnarray}
\label{eqn:optimal_C}
&\mathrm{maximize}& U^I_S=C\times b^s_{C+1}  \nonumber\\
&\mathrm{subject\ to}& b^s_{C+1}\ \textrm{is the $C+1$-th highest among $\widetilde{b}^i_{k}$,} \nonumber\\
&& (\ref{eqn:bid_channel}) \nonumber\\
&\mathrm{variables}& \{C\}.
\end{eqnarray}

However, there is no close-form solution for the problem (\ref{eqn:optimal_C}) since there is no close-form expression for (\ref{eqn:bid_channel}) in the general case.
The SH can leverage numerical method to obtain the optimal value of $C$.
From (\ref{eqn:guardband}), we have:
\begin{equation}
C\leq C_{MAX}=\left\lfloor \frac{B_0-b_0}{b_0}\right\rfloor.
\end{equation}
Therefore, an binary search can be perform within the range of $[0, C_{MAX}]$ for an optimal $C$.


\section{Economic Properties and Time Complexity}\label{sec:economic_prop}
In this section, we prove the properties of FlexAuc:
truthfulness, individual rationality, and efficiency.
We also analyze its time complexity.
The proofs of the following theorems are provided in the Appendix.

\subsection{Truthfulness}

\begin{theorem}
\label{th:truthful}
FlexAuc is truthful with any of the three payment mechanisms such
that $\mathcal{W}_i$'s best bidding strategy is $\{\widetilde{b}^i_{1}, \cdots, \widetilde{b}^i_{k},
\cdots, \widetilde{b}^i_{C}\}$.
\end{theorem}

\subsection{Individual rationality}

\begin{theorem}
\label{th:rational}
FlexAuc is individual rational. In another word, any $\mathcal{W}_i$ will not pay more than its true valuation. Mathematically, for any $\mathcal{W}_i$, $c^i_j\leq \widetilde{b}^i_{j}, \forall 1\leq j\leq K_i$.
\end{theorem}

Individual rationality is a necessary property to motivate participants in the auction.

\subsection{Efficiency}
\begin{theorem}
\label{th:welfare}
FlexAuc maximizes social welfare.
\end{theorem}

\subsection{Time Complexity}

We now analyze the running time of the algorithms of FlexAuc.

For Algorithm \ref{alg1}, building heap takes $O(N)$. Heap
adjustment takes $O(\log N)$. The complexity of Algorithm
\ref{alg1} is $O(C\log N)$. For Algorithm \ref{alg2}, lines $1$ to
$3$ take $O(C\log N)$. Lines $4$ to $17$ take $O(NC)$. So
the complexity of Algorithm \ref{alg2} is $O(CN)$. Both Algorithms
\ref{alg3} and \ref{alg4} take $O(N)$ time.

It means that FlexAuc with three payment mechanisms induces the
computational overhead $O(NC)$, $O(N+C\log N)$ and $O(N+C\log N)$
respectively. Since SH receives in total $NC$ bids from
WSPs, the overhead is definitely acceptable.

\section{Numerical Results}\label{sec:num}

Here we present simulation results to verify our theoretical analysis,
evaluate the performance and compare it with existing mechanisms.
The experiment environment is MATLAB.

We first present the WSPs' strategies in terms of bidding and service pricing.
Then we show properties of FlexAuc, including:
i) the truthfulness of FlexAuc, which is our key design target;
ii) the impact of payment mechanisms on the SH's revenue;
and iii) comparison between FlexAuc and a scheme from existing works.
Finally, we show the SH's strategies on spectrum partitioning and the impact of the size of the guard band.

\subsection{Settings}

The default settings of parameters are as follows.
$B_0=50$ (MHz). $b_0=0$ (MHz). $N=10$.
$N_i$ is randomly distributed within $[500, 1000]$ for
$i=1, \cdots, N$. $\{\alpha_i\}$s are equally distributed in
$[0.2, 0.4]$. The transmission range to the base station is randomly
chosen in $[500, 1000]$ (meter) for any user with a uniform
distribution. Assume that $75\%$ of users are indoor and their
service requirement is originated from indoor environment, and
$25\%$ from outdoor environment.

We define the attenuation factors as the multiplicative inverses of
the path-losses based on the ITU and COST models
\cite{recommendation19971225}.
\begin{enumerate}
\item From base station to outdoor user $U^i_{j}$:
\begin{equation}
H_j=10^{-4.9}(\frac{r}{1000})^{-4}f^{-3}10^{-\mu_0/10}; \notag
\end{equation}
\item From base station to indoor user $U^i_{j}$:
\begin{equation}
H_j=10^{-3.7}(\frac{r}{1000})^{-3}10^{-\mu_0/10}10^{-\frac{18.3n(\frac{n+2}{n+1}-0.46)}{10}};
\notag
\end{equation}
\end{enumerate}

The other default values of parameters are as follows: $P_i=1$
(watt), $n_0=-204$ (dB/Hz), $r$ (in meters) is the transmission
range, $f=2000$ (MHz) is the carrier frequency, $n=20$ is the number
of floors in the path, $\mu_0$ is the log-normal shadowing factor with
the standard deviation of $8$ (dB).

%

In order to compare our mechanism with previous auction mechanism, a frequently used
scheme that restricts each WSP to submit only one bid and the SH to make a
$0/1$ allocation \cite{zhou2009trust}\cite{feng2012tahes} is introduced as a baseline scheme.
We call it the OneBid auction.
In the OneBid, each $\mathcal{W}_i$ instead of submitting bids with flexible demands and valuations, the bid vector looks like $\{b^i_{1}>0, 0, \cdots, 0\}$.

All results in the following sections have been averaged over 100 cases with randomly generated parameters.

\subsection{Results}

\begin{figure}[t]
  \begin{minipage}{1.7in}
    \centering
    \includegraphics[width=1.7in]{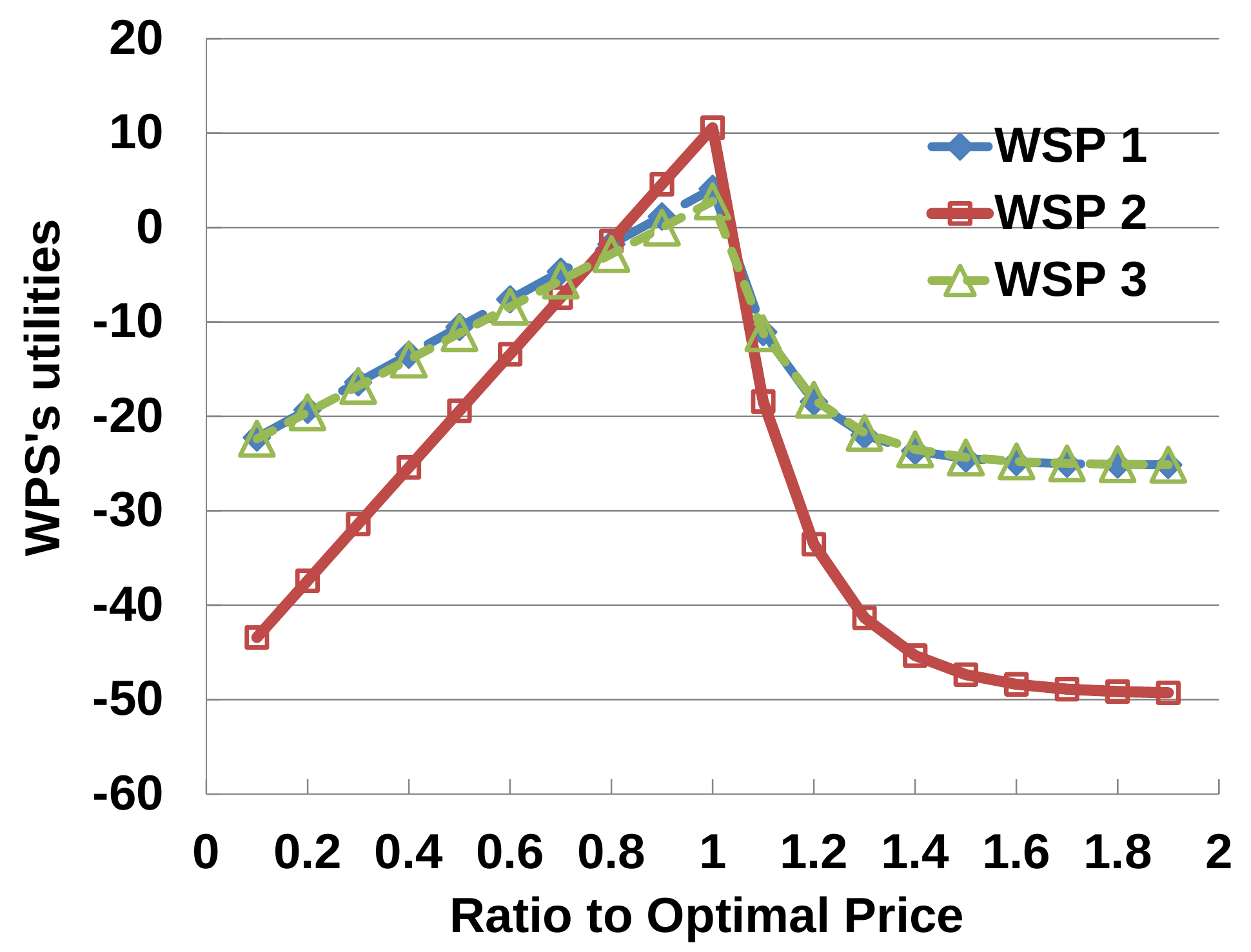}
    \caption{WSPs' optimal pricing strategies.}
    \label{fig:optprice}
  \end{minipage}
  \begin{minipage}{1.7in}
    \centering
    \includegraphics[width=1.7in]{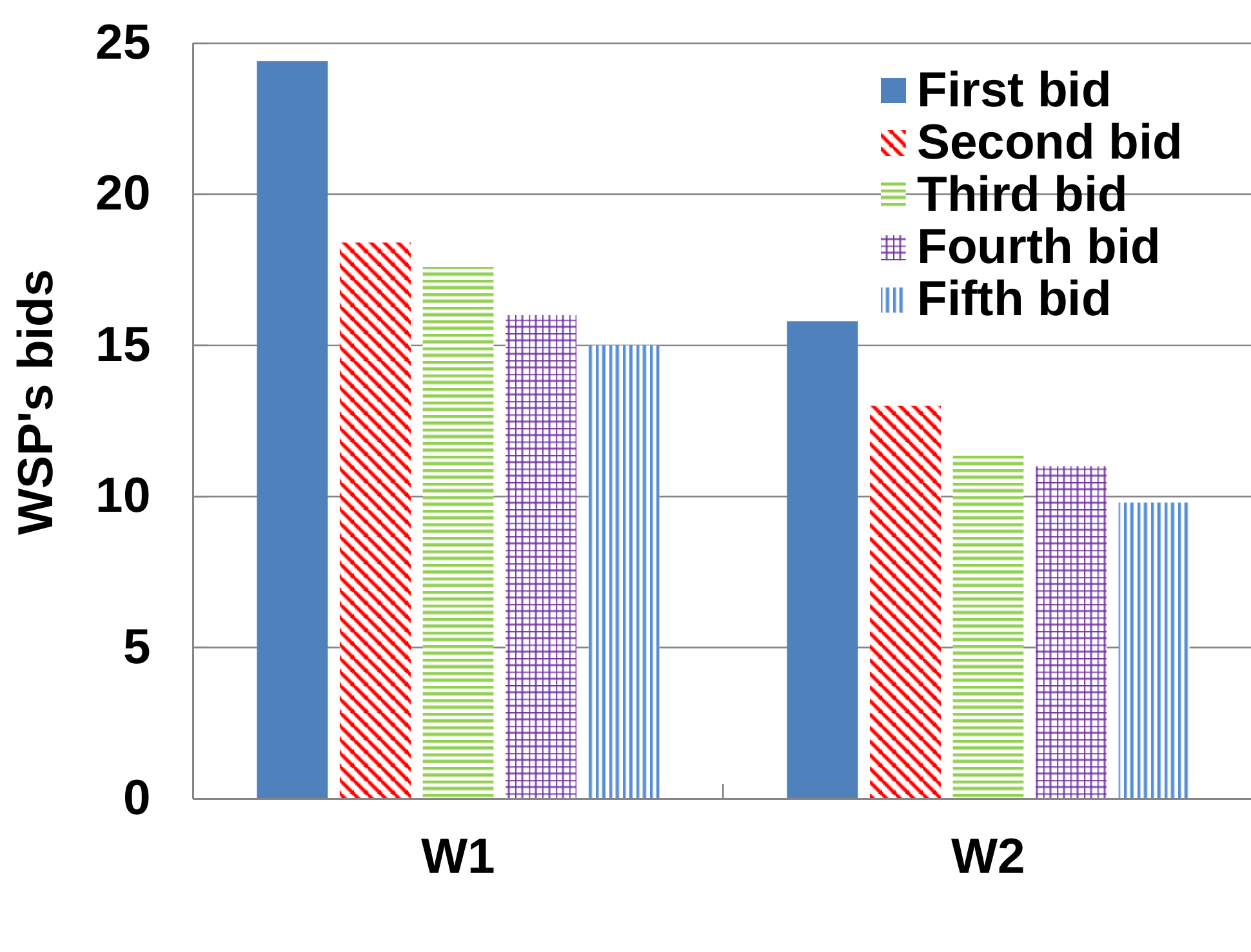}
    \caption{WSPs' bids structure follows the
    marginal effect.} \label{fig:wspbids}
  \end{minipage}
\end{figure}

\subsubsection{Strategies of the WSPs}

To observe the auction results and WSP's strategies, we assume the size of guard band is 0 and fix the number of channels for sale as $C=5$. Then there will be at most winners in the auction.

Fig. \ref{fig:optprice} shows the WSPs' utilities under different
pricing strategies. We select the three winning WSPs after the
auction and calculated their utilities under prices of different ratio of the optimal one:
$[0.1p^*_i, 0.2p^*_i, \ldots, 2p^*_i]$. We can see that setting higher or lower
prices (other than the ratio of 1) may lead to their overall losses. It verifies the correctness
of WSPs' optimal pricing strategies.

Fig. \ref{fig:wspbids} presents two WSPs' bids structure. The
results verify our theoretical analysis of the diminishing
marginal value of obtained channels (Theorem \ref{th:bid}). In this figure, $W_2$'s first
bid value is small than $W_1$'s fourth bid. If $W_2$ wins one
channel, then $W_1$ must win at least four channels.

\begin{table}[t]
\renewcommand{\arraystretch}{1.4}
\begin{center}
\caption{Truthfulness of WSPs' bids.} \label{tab:t4} \vspace{1ex}
\begin{tabular}{c c c c}
\hline $N$, $C$ &  $U_i^j(u)<U_i^j(t)$  & $U_i^j(u)=U_i^j(t)$ & $U_i^j(u)>U_i^j(t)$  \\
\hline
10, 5 & $0.1012$ & 0.8988 & 0 \\
10, 10 & $0.1331$ & 0.8669 & 0 \\
10, 20 & $0.1721$ & 0.8279 & 0 \\
\hline
\end{tabular}
\end{center}
\end{table}

\subsubsection{Truthfulness of FlexAuc}

In order to show the truthfulness of FlexAuc, we show the utility cannot be enhanced if the WSPs are not bidding truthfully.
We generate $100$ cases with random radio parameters and randomly
selected payment mechanism. In each case, we do $100$ times of
random selection of one WSP (either winner or loser) and make a
random adjustment from its true bidding values (while still keeping
the marginal effect of bids). Table \ref{tab:t4} shows the
truthfulness of the bids. $U_i(u)$ and $U_i(t)$ are WSP $i$'s
utilities when he bids untruthfully and truthfully respectively. The
values are the probabilities of the three cases with different $N$
and $C$ values. We see that in any case, $U_i^j(u)>U_i^j(t)$ never
happens, which supports the truthfulness of our auction (Theorem \ref{th:truthful}).


\begin{figure*}[t]
\centering
\subfloat[$C =3$.]
{\centering \includegraphics[width=1.6in]{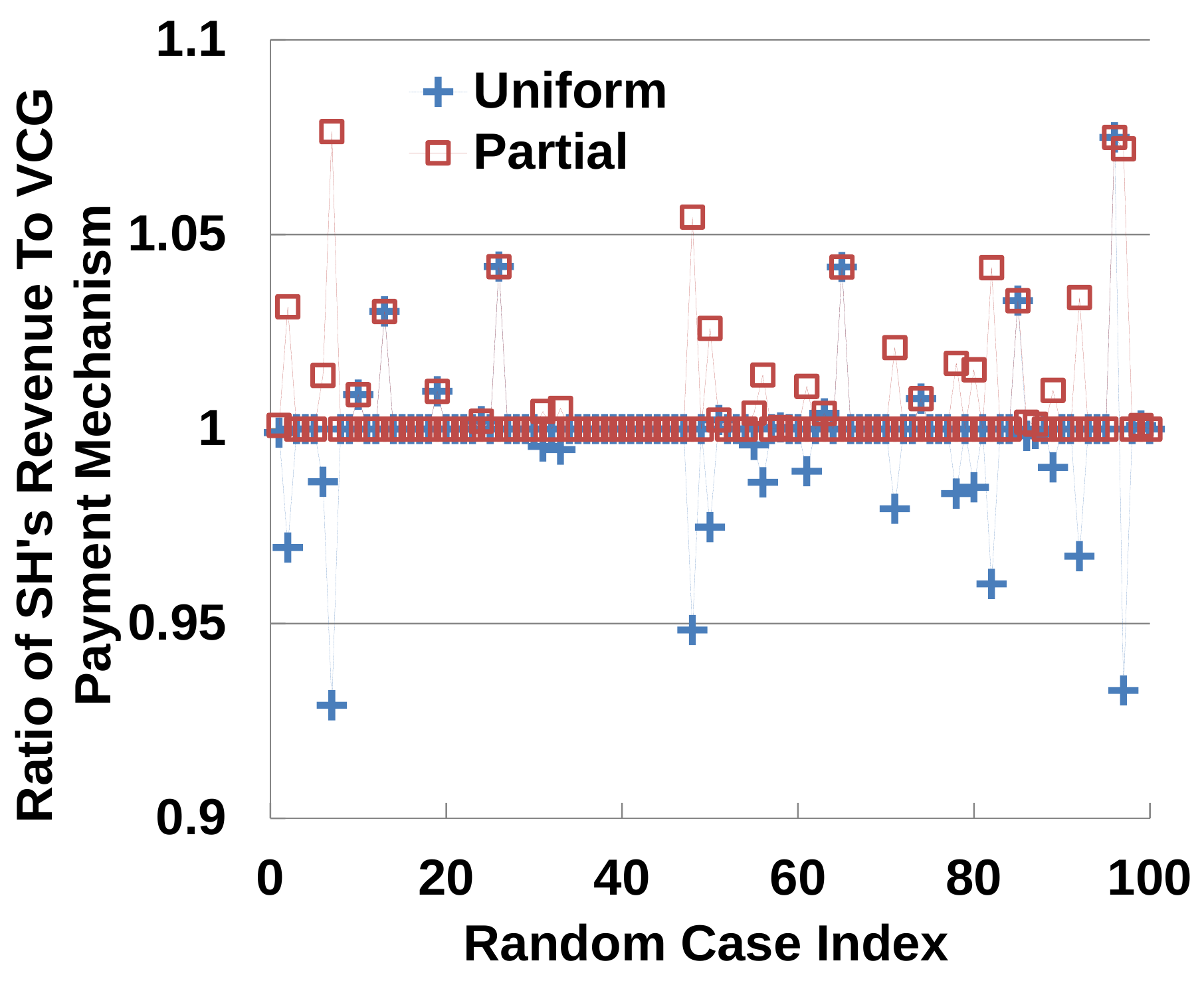}         
\label{fig:3channel}}%
\subfloat[$C =5$.]
{\centering\includegraphics[width=1.6in]{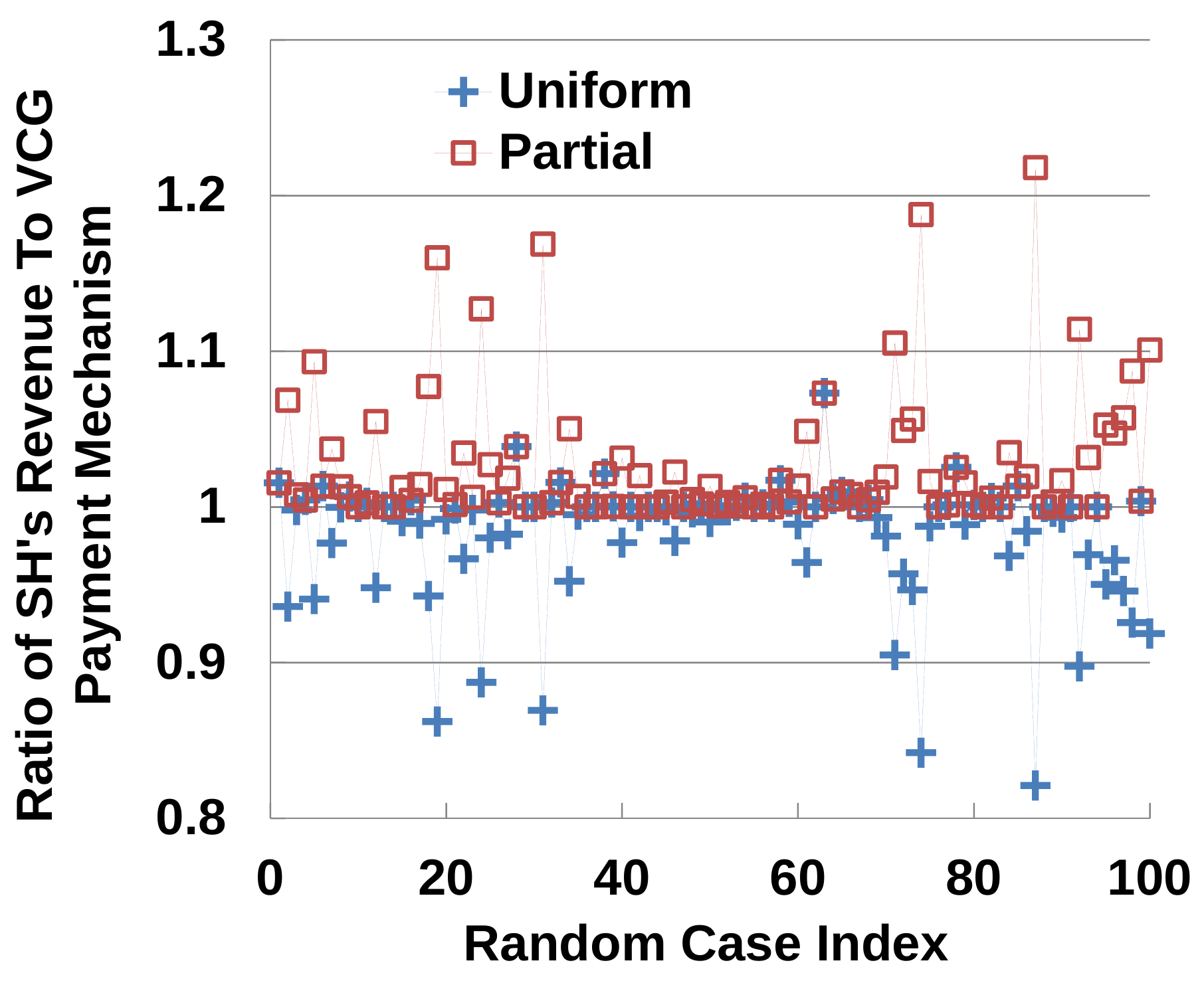}           
\label{fig:5channel}}%
\subfloat[$C =7$.]
{\centering\includegraphics[width=1.6in]{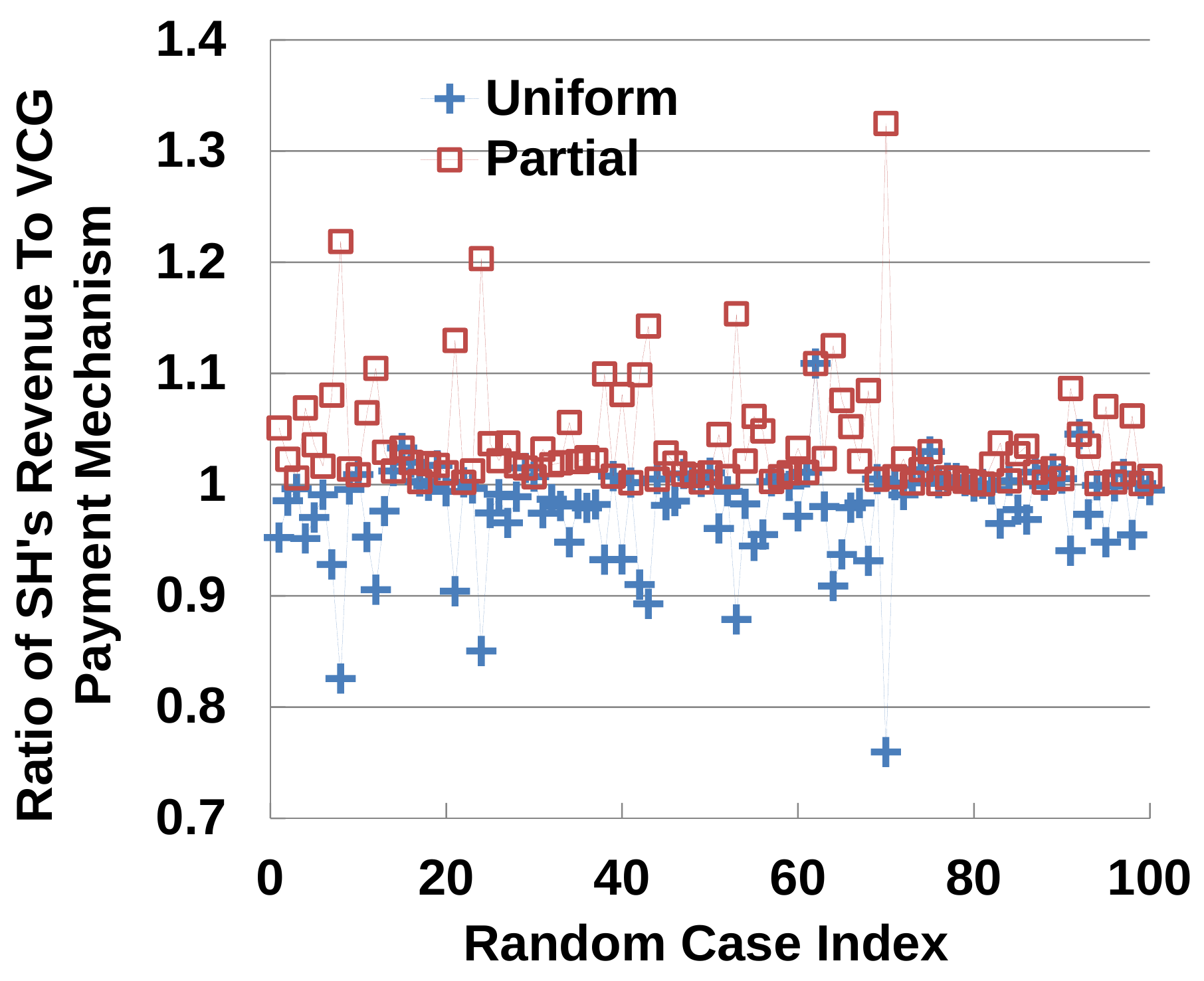}           
\label{fig:7channel}}%
\subfloat[$C>N$.]
{\centering\includegraphics[width=1.6in]{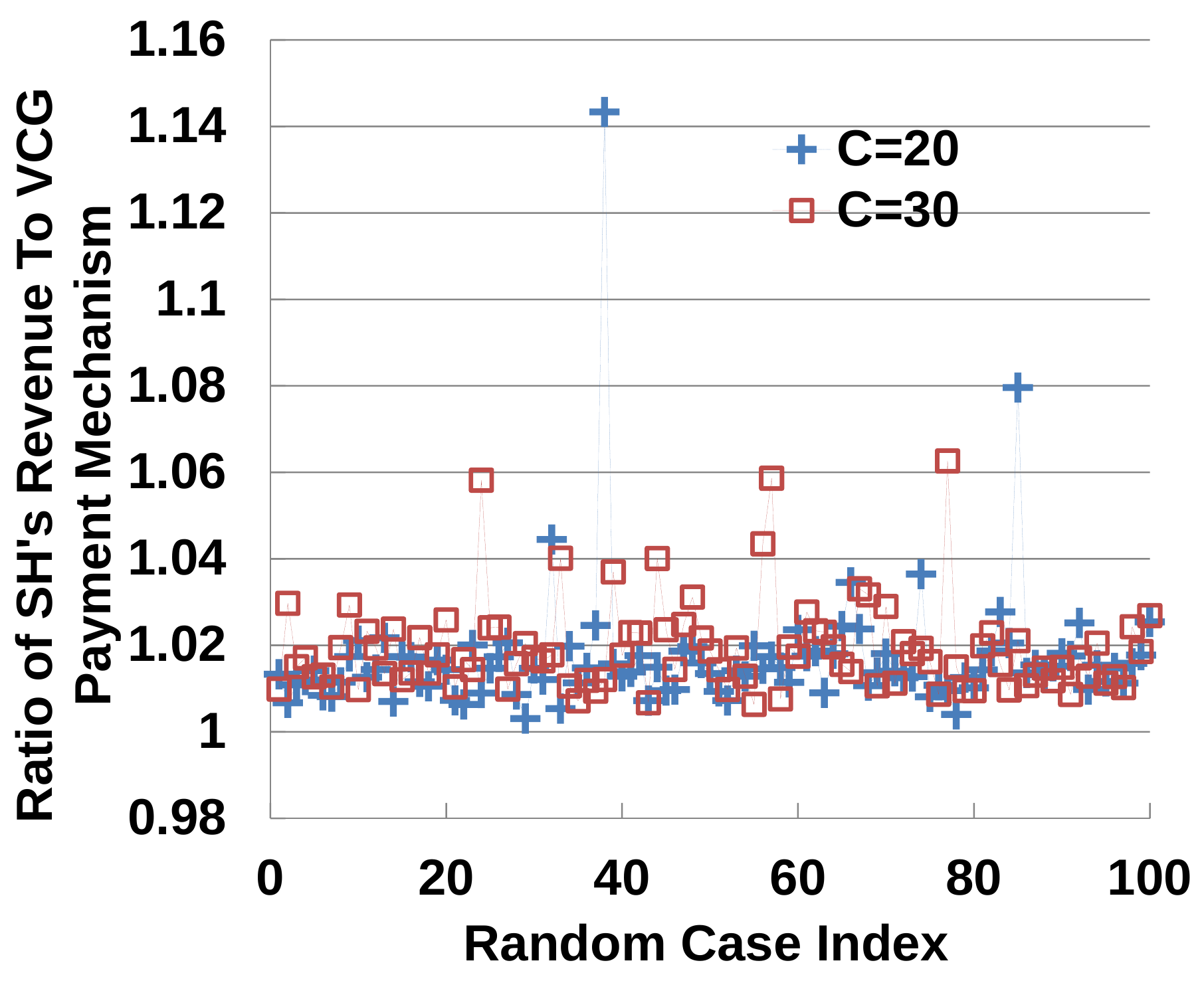}           
\label{fig:2030channel}}%
\caption{Comparison of three payment mechanisms.}
\label{fig:payment357}
\vspace{-0.0cm}
\end{figure*}

\subsubsection{Impact of FlexAuc's Payment Mechanisms}

To observe the impact of the payment mechanisms. We assume the size of guard band is 0 and varies $C$ in the range of $[3, 5, 7, 20, 30]$. In the first three cases, $C<N$, while $C>N$ in the last two cases. We calculate the revenue generated by the three payment mechanisms. We treat the VCG scheme as a benchmark and calculate the ratio of the revenues generated by the other two schemes and to those from the VCG scheme.

When $C<N$, Fig. \ref{fig:3channel}, \ref{fig:5channel} and \ref{fig:7channel}
compare the revenue from the three pricing mechanisms in the auction.
We observe that the partial uniform pricing always generate revenue for SH no worse than VCG (ratio $\geq$ 1) and uniform pricing, which verifies Lemma \ref{lemma:pricing}. However, the uniform pricing does not outperform VCG all the time and vice versa.
Also, the performance gaps of the three mechanisms increase with $C$. When the $10$ WSPs bid for $7$ channels, uniform pricing generates sub-optimal results for the SH in most cases.

When $C>N$, uniform pricing does not work any more. We compare the revenue from the
remaining two policies. In Fig. \ref{fig:2030channel}, all the data points are with value larger than 1, which verifies the advantages of partial uniform pricing under the two cases of $C=20$ and $C=30$.

\begin{figure}
\centering
\includegraphics[width=2.3in]{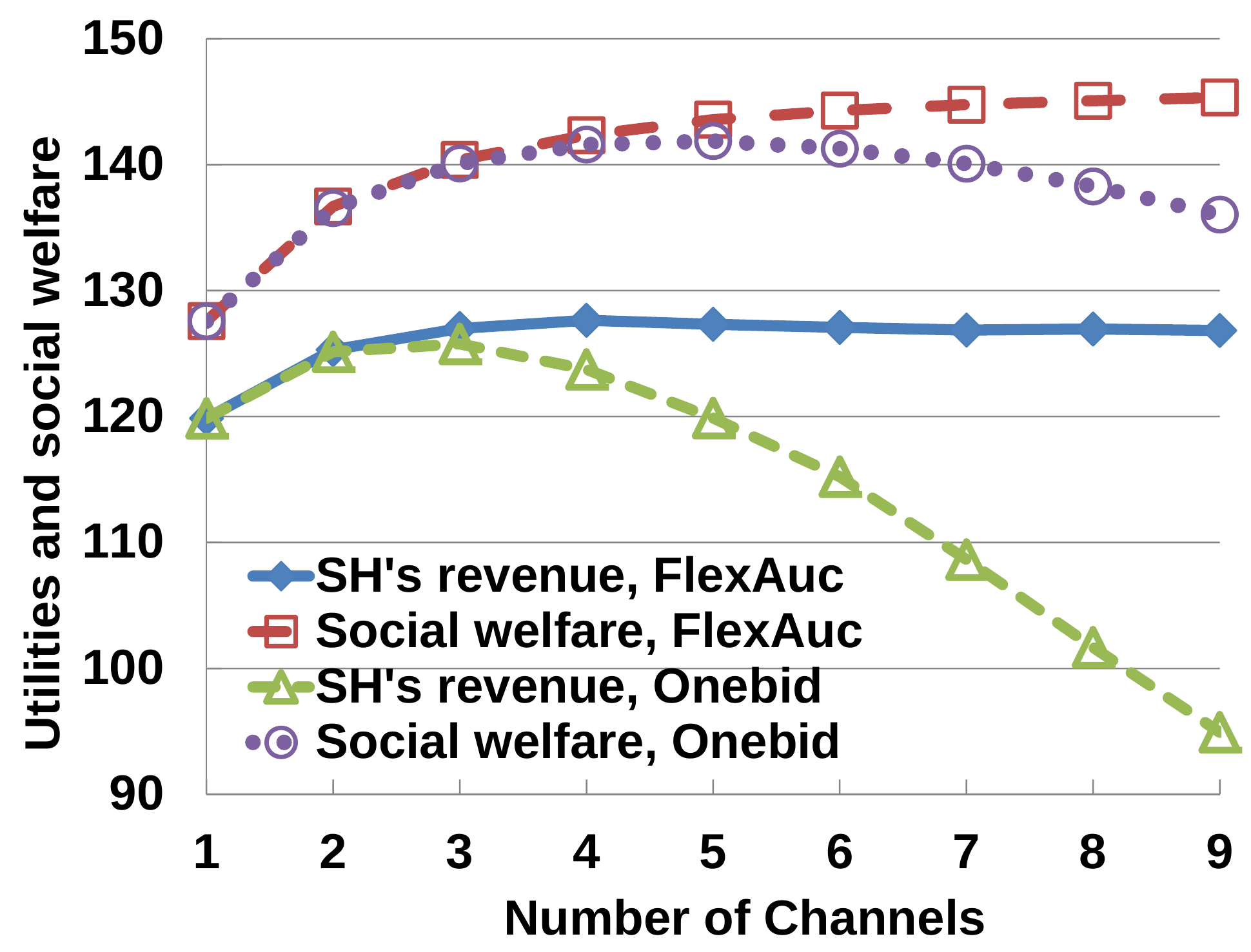}
\caption{Comparison of FlexAuc and OneBid mechanisms.}
\label{fig:onebid}
\end{figure}

\subsubsection{Comparison of FlexAuc and OneBid}

To show the advantage of FlexAuc in the aspect that it enables flexible demands and valuations,
we present the comparison between FlexAuc and the OneBid auction in Fig. \ref{fig:onebid}.
We set $C$ in the range of $[1, 2, \cdots, 9]$ and plot both the SH's revenue and the social welfare from the auction result.

We observe that when $C$ is small enough ($C \leq 3$), the two auction
almost perform the same in both SH's revenue and social welfare.
When $C$ is larger, FlexAuc outperforms the OneBid and the gaps
keep increasing in both revenue and social welfare.
By enabling the flexible auction, both the revenue of the SH and the social welfare can be greatly increased.

\begin{figure*}[t]
\centering
\subfloat[$b_0=0$.]
{\centering \includegraphics[width=1.6in]{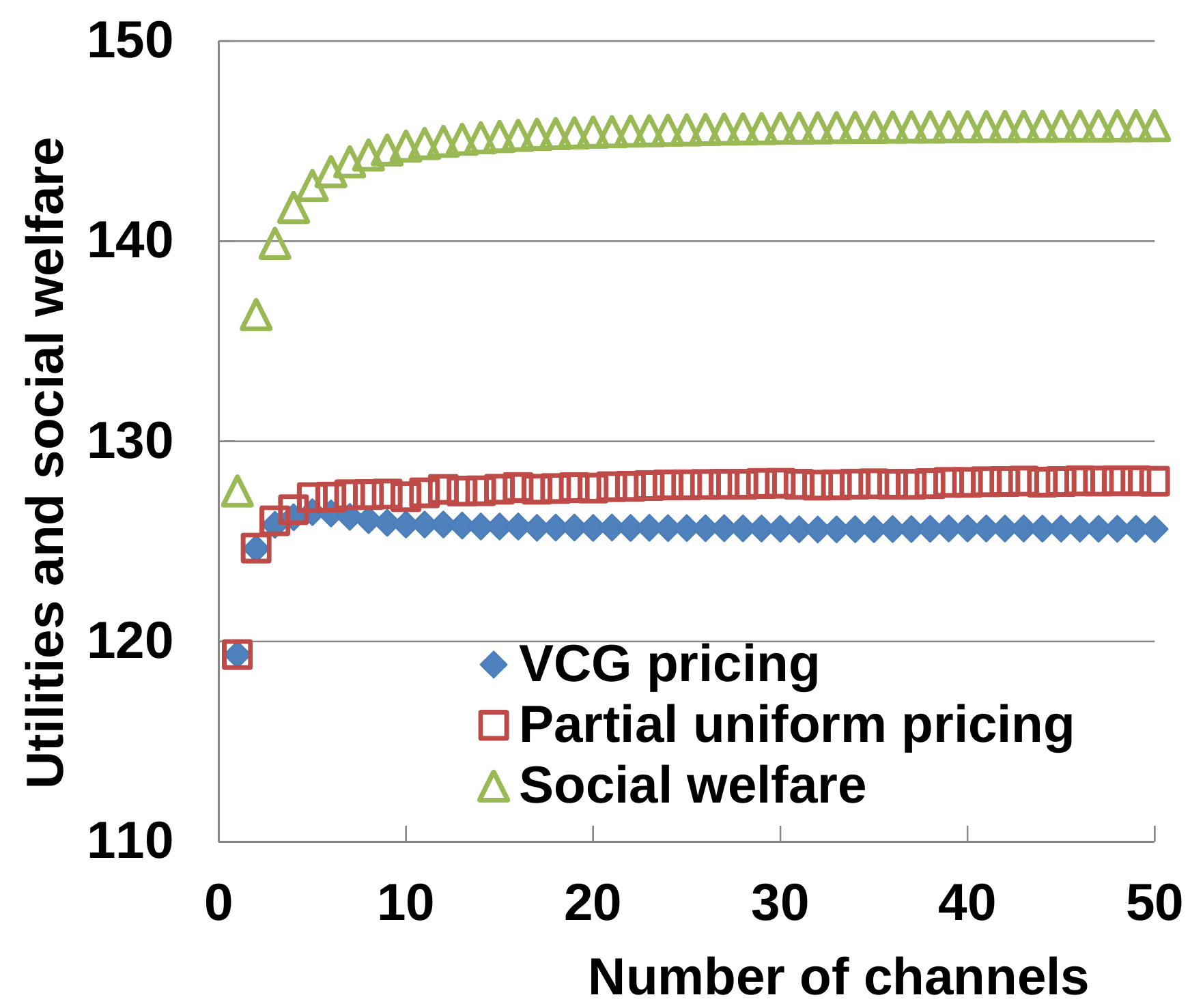}         
\label{fig:partitionb0}}%
\subfloat[$b_0=0.1$.]
{\centering\includegraphics[width=1.6in]{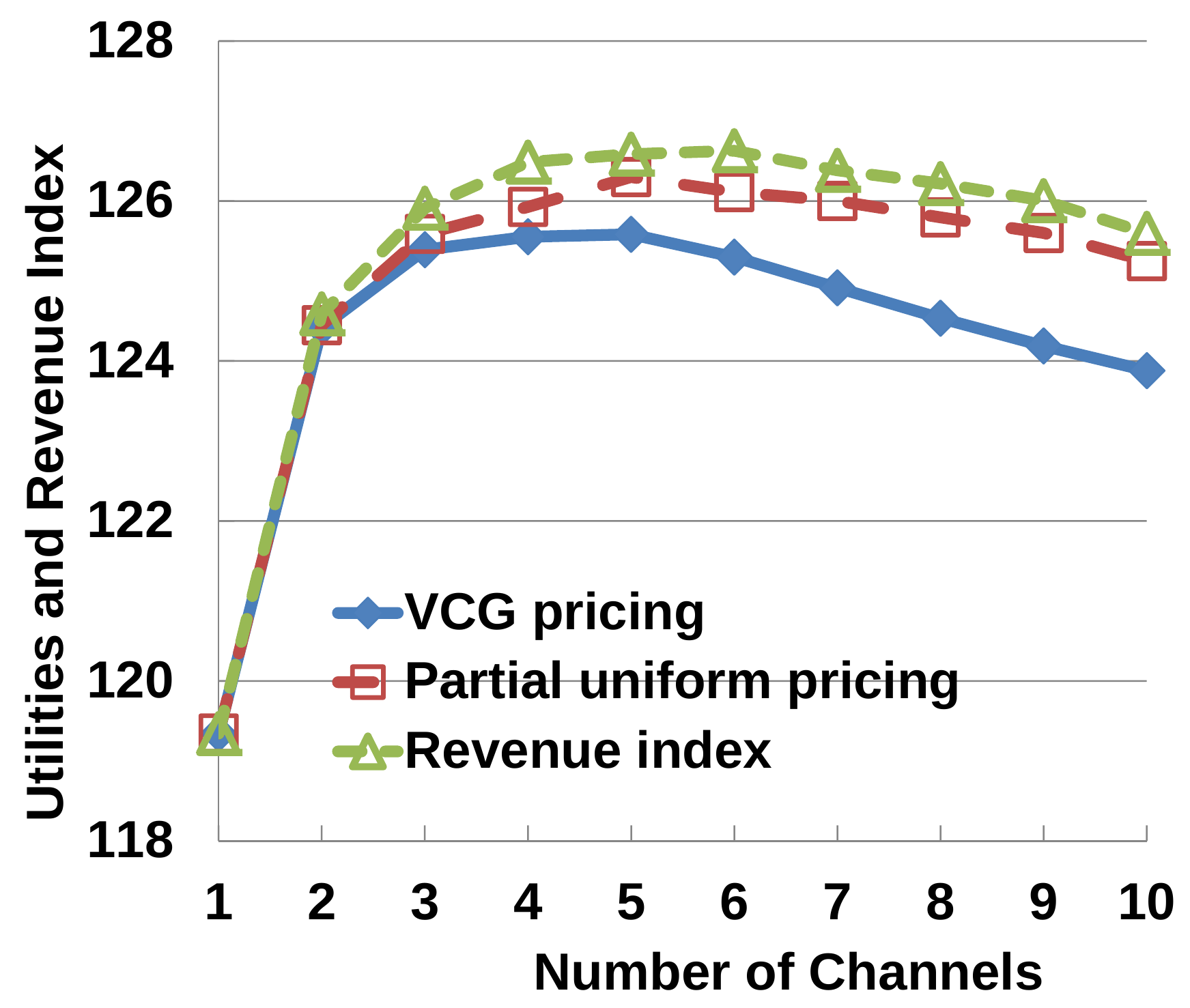}           
\label{fig:partitionb1}}%
\subfloat[$b_0=0.5$.]
{\centering\includegraphics[width=1.6in]{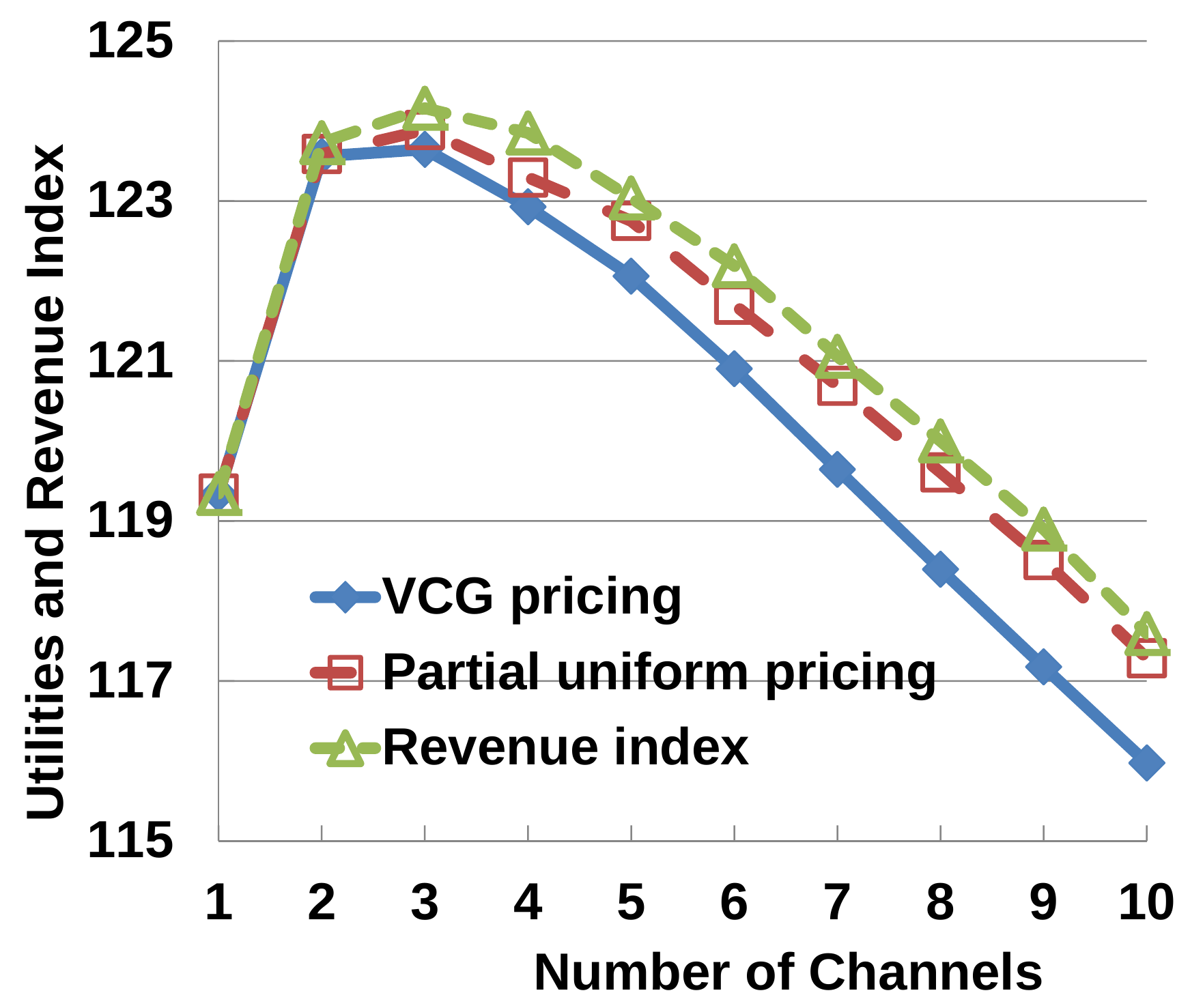}           
\label{fig:partitionb2}}%
\subfloat[$b_0=1$.]
{\centering\includegraphics[width=1.6in]{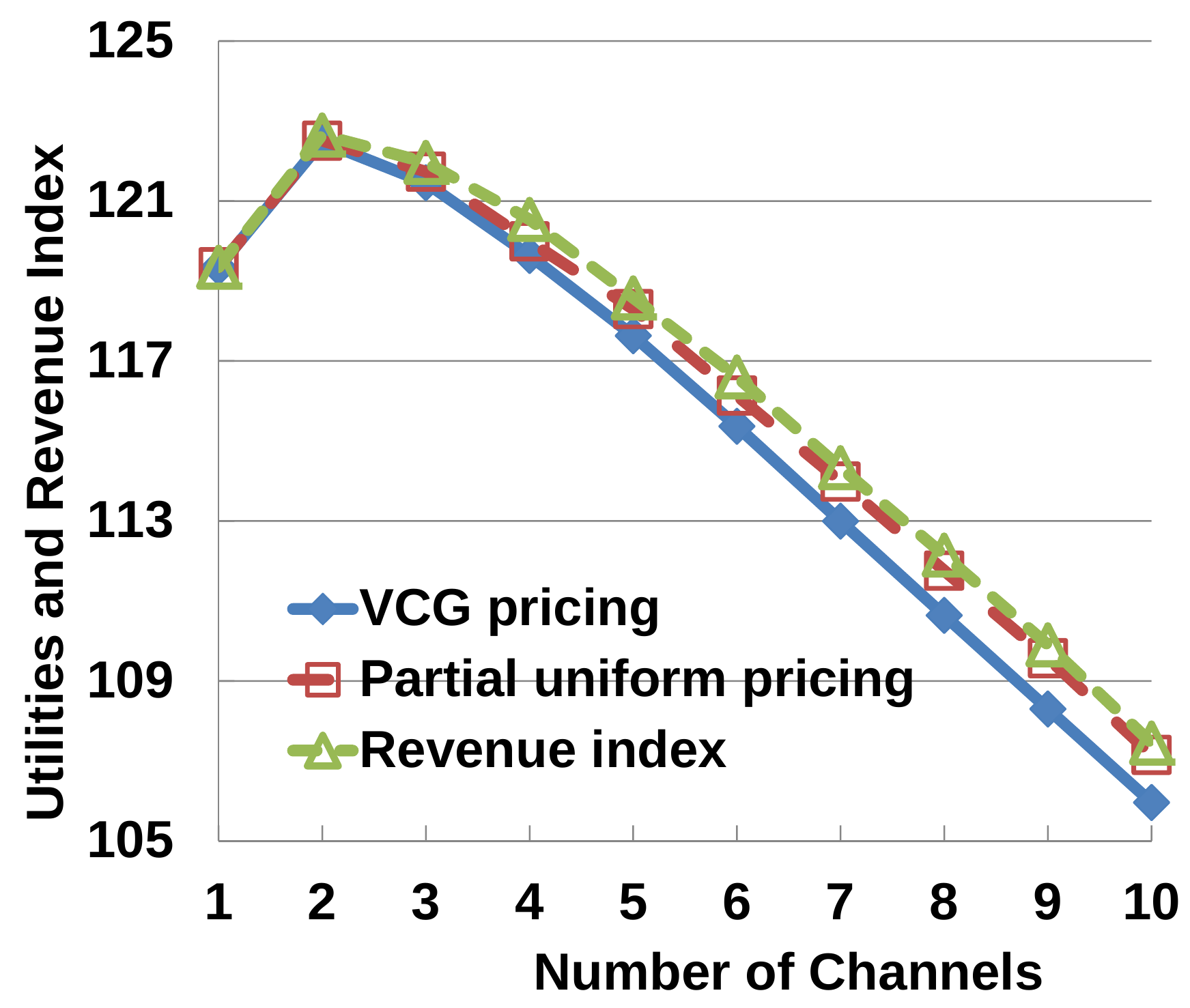}           
\label{fig:partitionb3}}%
\caption{Influences of $C$ on the SH's revenue.}
\label{fig:influencesofC}
\vspace{-0.0cm}
\end{figure*}

\subsubsection{Strategy of the SH}

We show in this part how the SH determine the optimal channelization scheme.
We vary the size of the guard band in the range $[0, 0.1, 0.5, 1]$ and plot the average revenue from the VCG and partial uniform pricing schemes under different $C$ in Fig. \ref{fig:influencesofC}. Since optimal $C$ may be larger than $N$, we omit the case of uniform pricing here.
We plot the social welfare for $b_0=0$ and the corresponding $U^I_S$s under other $b_0$s.

Fig. \ref{fig:partitionb0} gives numerical results for optimal $C$ when $b_0=0$. We see that the larger $C$ is, the higher social welfare it achieves. When $C$ is large enough, the increment of social welfare
gets smaller.
Strictly, we have the property that $S(kC) \geq S(C)$ under the same auction rule, where $k$ is any
positive integer. An intuition here is that after the further partition of current
channels, the SH can allocate $k$ times of number of channels to current
winning WSPs, which leads to the same social welfare. So the further
partition of channels does not decrease social welfare.

From Fig. \ref{fig:partitionb1},\ref{fig:partitionb2},\ref{fig:partitionb3}, we can see that when $b_0\neq 0$. There exist a $C$ that maximize the SH's revenue in either payments schemes. In all the three cases, $U^I_S$ is no less than the revenue provide by either payment schemes, which verifies that it is their upper bound (Lemma \ref{lemma:indicator}).


\subsubsection{Impact of the Guard Band}

From Fig. \ref{fig:influencesofC}, we can also observe the impact of the guard band on the SH's revenue and strategies. The results in the four sub-figures are based on the same random settings of $G_i$ and $\alpha_i$. First, we observe that the wider the guard band is, the $C$ resulting in the highest revenue will also be larger. When $b_0=0$, the highest revenue will be achieved with an infinite $C$.
Second, we also see that when $b_0$ is wider, the achieved revenue under the same $C$ would be smaller. That is due to the less available total bandwidth to sale in the auction with wider guard band.

\section{Conclusion}\label{sec:conclusion}

The WSPs face dynamic and diversed users' demands which impact their decisions on how much bandwidth to purchase and how much money to pay for the SH.
Previous spectrum auction studies do not pay attention to the tight relationship between bidding strategies and service provisions.
Existing auction mechanism cannot be directly applied in this scenario or with efficient computational performance.
In this paper, we analyze the end users' demands response, WSPs' optimal pricing and bidding strategies, SH's auction design and discuss the spectrum partition.
We propose FlexAuc as the solution framework.
FlexAuc consists of a standard winner determination part and a flexible payment mechanism.
There are three payment mechanisms studied and compared.
All of them are truthful and maximize social welfare.
The computational overhead of FlexAuc is linear to the input size of the bids.
We conduct comprehensive numerical simulation to verify our conclusions.

\section*{Appendix}

\subsection{Proof of Lemma \ref{lemma:indicator}}
\begin{proof}
For all the three payment mechanisms, unit payment of one channel related to one or more losing bids in the auction, for any $\widetilde{W}_i$ and $1\leq j\leq K_i$, $c^i_j\leq b^s_{C+1}$.
Moreover, $\sum_{i=1}^NK_i=C$.
Therefore,
\begin{equation}
\sum_{i=1}^N\sum_{j=1}^{K_i}\widetilde{b}^i_{j}\leq Cb^s_{C+1},
\end{equation}
The left of which equals to $U_S$.

The above bound is also tight. We can construct a case in which $b^s_{C+1}$ equals to $\widetilde{b}^z_{1}$. Therefore, the highest bid of the losing buyer $\widetilde{W}_z$ will be the unit price for all the three payment mechanisms.
\end{proof}

\subsection{Proof of Theorem \ref{th:bid}}
\begin{proof}
By Eq. (\ref{formula:1c})(\ref{formula:kc}), we have
\begin{equation}
\sum_{k=1}^{K_i} \widetilde{b}^i_{k}=p^*_i(K_i)\cdot
\min\left\{\sum_{j=1}^{N_i}w^{*i}_j(K_i), K_iB \right\} .
\label{formula:totaldemand}
\end{equation}
Substitute Eq. (\ref{formula:optimalprice}) into it, then we get
\begin{equation}
\sum_{k=1}^{K_i} \widetilde{b}^i_{k}=\min\left\{\left(\ln\frac{G_i}{K_iB}-1\right)K_iB,
G_ie^{-2}\right\}.
\end{equation}
Define function
\begin{equation}
f(K_i)=\left(\ln\frac{G_i}{K_iB}-1\right)K_iB
\end{equation}
Because
\begin{equation}
\frac{\partial f(K_i)}{\partial K_i}=B\left(\ln\frac{G_i}{K_iB}-2\right) \geq
0
\end{equation}
and
\begin{equation}
\frac{\partial^2 f(K_i)}{\partial K_i^2}=-\frac{B}{K_i}< 0.
\end{equation}
$G_ie^{-2}$ is a constant and the operator $\min$ keeps the
convexity. So Eq. (\ref{formula:totaldemand}) is a convex function.

Note that $f(K_i)$ is positive for all $K_i=1,2,\cdots$. By
$f(K_i)$'s continuity, we have
\begin{equation}
\begin{aligned}
f(0)&=\lim_{K_i \rightarrow 0^+}f(K_i)
=\lim_{x=\frac{1}{K_i} \rightarrow +\infty}
\frac{B\left(\ln\frac{GX}{B}-1\right)}{x} \\
&=\lim_{x \rightarrow +\infty}\frac{B}{x} =0.
\end{aligned}
\end{equation}

By Eq. (\ref{formula:totaldemand})'s convexity, we have
\begin{equation}
2\sum_{k=1}^{K_i+1} \widetilde{b}^i_{k} \geq \sum_{k=1}^{K_i}
\widetilde{b}^i_{k}+\sum_{k=1}^{K_i+2} \widetilde{b}^i_{k}
\end{equation}
for all $K_i=0,1,2,\cdots$. After simplification, we obtain
$\widetilde{b}^i_{1} \geq \widetilde{b}^i_{2} \geq \cdots \geq \widetilde{b}^i_{C}$.

\end{proof}

\subsection{Proof of Theorem \ref{th:truthful}}
\begin{proof}
To prove the truthfulness, we need to show that for any $1 \leq i
\leq N$ and $1 \leq k \leq C$, $\{\widetilde{b}^i_{1}, \cdots, \widetilde{b}^i_{k},
\cdots, \widetilde{b}^i_{C}\}$ weakly dominates any other $\{b^i_{1}, \cdots,
b^i_{k}, \cdots, b^i_{C}\}$.

First we show that by replacing only $b^i_{k}$ with $\widetilde{b}^i_{k}$ (if
$b^i_{k-1} \leq \widetilde{b}^i_{k} \leq b^i_{k+1}$), the new strategy
$b^i_{1}, \cdots, \widetilde{b}^i_{k}, \cdots, b^i_{C}$ weakly dominates the
original strategy $b^i_{1}, \cdots, b^i_{k}, \cdots, b^i_{C}$. By
original strategy $\mathcal{W}_i$ wins $K_i$ channels and by new one he wins
$K^*_i$ channels. We discuss the three possible cases:
\begin{enumerate}
\item Case $1$: $K_i=K^*_i$. $\mathcal{W}_i$ wins the same number of channels by
both strategies. By any of the three mechanisms, its payment is
determined by other WSPs' bids. So $\mathcal{W}_i$ utilities are the same by
both strategies.

\item Case $2$: $K_i<K^*_i$. It means $\mathcal{W}_i$ wins more channel by new
strategy. Its payment will be no more than the marginal benefit of
the additional channel(s). So its utility is improved or keep the
same by the new strategy. New strategy dominates original one.

\item Case $3$: $K_i>K^*_i$. By new strategies, $\mathcal{W}_i$ wins less
channels. $\mathcal{W}_i$'s bid on $k$-th channel does matter. $b^i_{k}$ is one
of the highest $C$ biddings but $\widetilde{b}^i_{k}$ is not. As both $b^i_{k}$
and $\widetilde{b}^i_{k}$ are larger than $b^i_{j}$ ($j=k+1,k+2,\cdots,C$), so
$b^i_{j}$ is not one of the highest $C$ biddings. So $K_i=K^*_i+1$.
That means the only difference is that original strategy gets $k$-th
channel and new strategy does not. According to the payment
mechanisms, by original strategy, $\mathcal{W}_i$ pays more than the marginal
benefit $\widetilde{b}^i_{k}$ to get the $k$-th channel. So New strategy
dominates original one.
\end{enumerate}

Then for any strategy $\{b^i_{1}, \cdots, b^i_{C}\}$, we can adjust it
in reverse order repeatedly like bubble sort algorithm: $\{b^i_{1},
\cdots, b^i_{C}\} \rightarrow \{b^i_{1}, \cdots,
\min\{\widetilde{b}^i_{C},b^i_{C-1}\} \} \rightarrow \{b^i_{1}, \cdots,
\min\{\widetilde{b}^i_{k},b^i_{k-1}\}, \cdots \} \rightarrow \cdots
\rightarrow \{\widetilde{b}^i_{1, \cdots}\} \rightarrow \cdots \rightarrow
\{\widetilde{b}^i_{1}, \widetilde{b}^i_{2} \cdots \} \rightarrow \cdots$ until it becomes
$\{\widetilde{b}^i_{1}, \cdots, \widetilde{b}^i_{C}\}$. It can be achieved by no more than
$\frac{C(C+1)}{2}$ adjustments. During the adjustments, the new
strategies dominant the old ones. So $\{\widetilde{b}^i_{1}, \cdots,
\widetilde{b}^i_{C}\}$ dominates any $\{b^i_{1}, \cdots, b^i_{C}\}$.
\end{proof}

\subsection{Proof of Theorem \ref{th:rational}}
\begin{proof}
Suppose $\widetilde{b}^i_{j}$ is a winning bid from $\widetilde{W}_i$ and $c^i_{j}$ is the corresponding price.
According to the winner selection procedure and the payment mechanisms, we have $\widetilde{b}^i_{j}\geq b^s_C$ and $c^i_{j}\leq b^s_{C+1}$.
Furthermore, $\{b^s_{k}\}$ are sorted in descending order in the auction, we have $b^s_C\geq b^s_{C+1}$.
Therefore, $\widetilde{b}^i_{j}\geq c^i_{j}$ holds.
\end{proof}

\subsection{Proof of Theorem \ref{th:welfare}}
It is well-known that VCG mechanism maximizes social welfare. We
provide a simple proof here.

\begin{proof}
An auction rule $\mathbf{R^*}$ is efficient if it maximizes social
welfare,
\begin{equation}
\mathbf{R^*}(\mathbf{x}) \in arg \max_{\forall \mathbf{R}} \sum
E_jx_j,
\end{equation}
where $x_j={0,1}$. By VCG mechanism, the payment of bidder $i$ is
\begin{equation}
c_i=W(0,\mathbf{x}_{-i})-W_{-i}(\mathbf{x}).
\end{equation}
Bidder $i$'s utility is
\begin{equation}
\begin{aligned}
E_i-c_i &= E_i+W_{-i}(\mathbf{x})-W(0,\mathbf{x}_{-i})\\
&=W(\mathbf{x})-W(0,\mathbf{x}_{-i}).
\end{aligned}
\end{equation}
So maximization of its own utility is equivalent to maximization of
social welfare.

We know that the uniform pricing auction and partial uniform pricing
auction distinguish from VCG auction only in the payment part. The
payment effect can be canceled out by the summation of WSPs' and
SH's utilities. So auctions with the two payment mechanism also
maximize social welfare. It means an auction maximizes social
welfare as long as it allocates items to those who evaluate them
most. By motivating WSPs to bid truthfully and selecting the highest
$C$ bids, FlexAuc indeed maximizes social welfare.
\end{proof}


\bibliographystyle{./IEEEtran}
\bibliography{./IEEEabrv,./flexauc-twc}

\end{document}